\iftrue
\documentclass[aps,prl,twocolumn,
			   groupedaddress,superscriptaddress,
			   amsfonts,amssymb,amsmath,
			   citeautoscript,
			   a4paper]{revtex4-1}
\else
\documentclass[aps,pra,preprint,
			   groupedaddress,superscriptaddress,
			   amsfonts,amssymb,amsmath,
			   citeautoscript,
			   a4paper]{revtex4-1}
\fi

\usepackage[pdftex]{hyperref}
\hypersetup{colorlinks,
			linkcolor={blue!75!black!80!yellow},
			citecolor={blue!75!black!80!yellow},
			urlcolor={blue!75!black!80!yellow},
			pdfstartview=FitH}
\usepackage{graphicx}
\usepackage{mathrsfs}
\usepackage{xspace}
\usepackage{braket}
\usepackage{xr}
\usepackage{gensymb} 
\usepackage{xcite}
\usepackage{xcolor,soul}
\usepackage{stmaryrd} 
\usepackage[UKenglish]{babel}
\usepackage{placeins} 

\usepackage{siunitx}
\DeclareSIUnit[number-unit-product=]\percent{\char`\%} 

\usepackage{txfonts}  
\usepackage{txfontsb} 

\makeatletter
\newcommand*{\addFileDependency}[1]{
  \typeout{(#1)}
  \@addtofilelist{#1}
  \IfFileExists{#1}{}{\typeout{No file #1.}}
}
\makeatother

\newcommand*{\myexternaldocument}[1]{%
    \externaldocument{#1}%
    \addFileDependency{#1.tex}%
    \addFileDependency{#1.aux}%
}

\makeatletter
\renewcommand\@make@capt@title[2]{%
	\@ifx@empty\float@link{\@firstofone}{\expandafter\href\expandafter{\float@link}}%
	\sffamily{\textbf{#1}}\@caption@fignum@sep#2
}%

\makeatother

\newcommand*\diff{\mathop{}\mathrm{d}}

\newcommand{\iu}{\mathrm{i}}
\newcommand{\e}{\mathrm{e}}

\newcommand{\appropto}{\mathrel{\vcenter{
			\offinterlineskip\halign{\hfil$##$\cr
				\propto\cr\noalign{\kern2pt}\sim\cr\noalign{\kern-2pt}}}}}

\newcommand{\hspin}{\ket{\text{h}}}
\newcommand{\vspin}{\ket{\text{v}}}
\newcommand{\pe}{\'{e}\xspace}
\newcommand{\T}{$\mathcal{T}$}

\newcommand{\matr}[1]{\boldsymbol{#1}}     
\newcommand{\sigmax}{\matr{\sigma}_x}
\newcommand{\sigmay}{\matr{\sigma}_y}
\newcommand{\sigmaz}{\matr{\sigma}_z}

\usepackage{textcomp} 
\usepackage{xifthen}
\usepackage{etoolbox}
\newboolean{togglecomments}
\newboolean{togglechanges}

\setboolean{togglecomments}{false}
\setboolean{togglechanges}{true}

\newcommand{\comment}[2]{%
    \ifbool{togglecomments}%
    {\textcolor{blue!70!black}{\small\textsf{%
    \textsuperscript{\textsc{\textsf{\MakeLowercase{#1}}}}%
    [#2]}}} 
    {}}     
\newcommand{\swap}[2]{\ifbool{togglechanges}
    {#2}  
    {\textcolor{red!70!black}{[#1]}\textrightarrow{}\textcolor{green!50!black}{[#2]}}}
\newcommand{\remove}[1]{\ifbool{togglechanges}
    {}    
    {\textcolor{red!70!black}{#1}}}
\newcommand{\inset}[1]{\ifbool{togglechanges}
    {#1}  
    {\textcolor{green!50!black}{#1}}}
\newcommand{\optional}[1]{\ifbool{togglechanges}
    {}    
    {\textcolor{yellow!50!orange!80!gray}{#1}}}

\newcommand{\citeremind}[1]{%
    [\textcolor{blue!75!black!80!yellow}{
        $\blacksquare$%
	    \ifthenelse{\isempty{#1}}
	        {}
	        {\textsuperscript{\tiny\textsf{#1}}}%
	}]\xspace}

\newcommand{\ie}{i.e.\@\xspace}  
\newcommand{\cf}{cf.\@\xspace}
\newcommand{\eg}{e.g.\@\xspace}

\newcommand{\mitaffil}{\footnotesize Research Laboratory of Electronics, Massachusetts Institute of Technology, Cambridge, Massachusetts 02139, USA}
\newcommand{\uppenaffil}{\footnotesize Department of Physics and Astronomy, University of Pennsylvania, Philadelphia, Pennsylvania 19104, USA}
\newcommand{\pkuaffil}{\footnotesize State Key Laboratory of Advanced Optical Communication Systems and Networks, Peking University, Beijing 100871, China}
\myexternaldocument{SI/SI}

\begin{document}

\title{
Synthesis and Observation of Non-Abelian Gauge Fields in Real Space
    }

\author{Yi~Yang}
\email{yiy@mit.edu}
\affiliation{\mitaffil}
\author{Chao~Peng}
\affiliation{\pkuaffil}
\author{Di~Zhu}	
\affiliation{\mitaffil}
\author{Hrvoje Buljan}
\affiliation{\footnotesize Department of Physics, Faculty of Science, University of Zagreb, Bijeni\v{c}ka c. 32, 10000 Zagreb, Croatia}
\affiliation{\footnotesize The MOE Key Laboratory of Weak-Light Nonlinear Photonics, TEDA Applied Physics Institute and School of Physics, Nankai University, Tianjin 300457, China}
\author{John~D.~Joannopoulos} 	\affiliation{\mitaffil}
\author{Bo Zhen}
\email{bozhen@sas.upenn.edu}
\affiliation{\uppenaffil}
\author{Marin~Solja\v{c}i\'{c}} \affiliation{\mitaffil}

\begin{abstract}
Gauge fields, real or synthetic, are crucial for understanding and manipulation of physical systems. The associated geometric phases can be measured, for example, from the Aharonov--Bohm interference. So far, real-space realizations of gauge fields have been limited to Abelian (commutative) ones. Here we report an experimental synthesis of non-Abelian gauge fields in real space and the observation of the non-Abelian Aharonov--Bohm effect with classical waves and classical fluxes. Based on optical mode degeneracy, we break time-reversal symmetry in different manners---via temporal modulation and the Faraday effect---to synthesize tunable non-Abelian gauge fields. The Sagnac interference of two final states, obtained by reversely-ordered path integrals, demonstrates the non-commutativity of the gauge fields. Our work introduces real-space building blocks for non-Abelian gauge fields, relevant for classical and quantum exotic topological phenomena.
\end{abstract}

\maketitle

Gauge fields are the backbone of gauge theories, the earliest example of which is classical electrodynamics. However, until the seminal Aharonov--Bohm effect~\cite{aharonov1959significance}, the scalar and vector potentials of electromagnetic fields have been considered as a convenient mathematical aid, rather than objects carrying physical consequences. It has been realized by Berry~\cite{berry1984quantal} that the Aharonov--Bohm phase imprinted on electrons can be interpreted as a real-space example of geometric phases~\cite{berry1984quantal,pancharatnam1956generalized}, which in fact appear in versatile physical systems.
For charge-neutral particles, such as photons~\cite{sounas2017non,yuan2018synthetic} and cold atoms~\cite{dalibard2011colloquium,eckardt2017colloquium,goldman2014light}, synthetic gauge fields can be created in real, momentum, or synthetic (\ie other parameters besides position or momentum) space.
These synthetic gauge fields enable engineered, artificial magnetic fields in systems of either broken or invariant time-reversal symmetry; and thus play a pivotal role in the realizations of topological phases~\cite{goldman2014light,lu2014topological,ozawa2018topological,aidelsburger2018artificial}, quantum simulations~\cite{bloch2012quantum,aspuru2012photonic}, and optoelectronic applications~\cite{tzuang2014non,fang2017generalized}.

Gauge fields are classified into Abelian (commutative) and non-Abelian (non-commutative), depending on the commutativity of the underlying group.
Synthetic Abelian gauge fields have been realized in various platforms including cold atoms~\cite{lin2009synthetic,aidelsburger2011experimental,miyake2013realizing,struck2012tunable,aidelsburger2013realization,aidelsburger2015measuring,jotzu2014experimental,ray2014observation,li2016bloch}, photons~\cite{fang2012realizing,fang2012photonic,hafezi2011robust,umucalilar2011artificial,li2014photonic,mittal2014topologically,schine2016synthetic,rechtsman2013strain,haldane2008possible,wang2009observation}, phonons~\cite{xiao2015synthetic,yang2017strain,abbaszadeh2017sonic}, polaritons~\cite{lim2017electrically}, and superconducting qubits~\cite{schroer2014measuring,roushan2014observation,roushan2017chiral}.
The synthesis of non-Abelian gauge fields is more challenging, due to the requirements of degeneracy and non-commutative, matrix-valued gauge potentials.
So far, they have been achieved only in the momentum and synthetic spaces.
Specifically, non-Abelian gauge fields have been realized in the momentum space using two-dimensional spin-orbit coupling~\cite{huang2016experimental,wu2016realization} in cold atoms.
In the synthetic space, non-Abelian geometric phases~\cite{wilczek1984appearance,wilczek1989geometric}, initially observed in nuclear magnetic resonances~\cite{zwanziger1990non,zee1988non,mead1987molecular,mead1992geometric}, have enabled non-Abelian geometric gates~\cite{abdumalikov2013experimental} and the simulation of an atomic Yang monopole~\cite{sugawa2018second}.
As yet, however, the realization of non-Abelian gauge fields in real space remains a bottleneck; therefore, the non-Abelian generalization of the Aharonov--Bohm effect---a real-space phenomenon that has stimulated longstanding theoretical interests~\cite{wu1975concept,horvathy1986non,alford1990discrete,chaichian2002aharonov,osterloh2005cold,jacob2007cold,iadecola2016non,chen2018non,dalibard2011colloquium,bohm2013geometric}---remains experimentally elusive.

Here we report the observation of the non-Abelian Aharonov--Bohm effect by synthesizing non-Abelian gauge fields in real space.
Exploiting a degeneracy in photonic modes, we create non-Abelian gauge fields by cascading multiple non-reciprocal optical elements that break the time-reversal symmetry (\T) in orthogonal bases of Hilbert space.
We demonstrate the genuine non-Abelian condition of our gauge fields in a fiber-optic Sagnac interferometer. The observed interference patterns show the signature features of the non-commutativity between a pair of time-reversed, cyclic evolution operators.
We also demonstrate that our synthetic magnetic fluxes are fully tunable, enabling controlled transitions between the Abelian and the non-Abelian regimes.
Taken together, our results lay the groundwork for the synthesis of non-Abelian gauge fields in real space, which provides basic ingredients for studying relevant single- or many-body topological states in photonic platforms.

Synthetic non-Abelian gauge fields demand a degeneracy of levels, which can, for example, be achieved by utilizing the internal degrees of freedom in quantum gases or exploiting the polarization/mode degeneracy and electromagnetic duality in photons.
For a particle moving along a closed path in a non-Abelian gauge field, its evolution operator reads
$
    \matr{W} \equiv \mathcal{P}\exp \iu \oint \matr{A}\diff\mathbf{l},
$
where $\mathcal{P}$ represents path-ordered integral and $\matr{A}$ is the matrix-valued gauge field.
Its trace, $W \equiv \mathrm{Tr}\,\matr{W}$, is gauge-invariant, and is also known as the Wilson loop~\cite{wilson1974confinement}.
For particles with N-fold degeneracies, the non-Abelian gauge fields can take forms of U(N). Here we focus on the SU(2) gauge fields, since our photonic system enables the definition of a pseudospin---a two-fold degeneracy in the polarization states.
Crucially, we focus on the situations where the involved gauge fields break \T-symmetry and state transport becomes nonreciprocal.
In what follows we illustrate the consequence of real-space guage fields on the pseudospin evolution in Hilbert space [\ie the Poincar\pe (or Bloch) sphere].

\begin{figure}[htbp]
	\centering
	\includegraphics[width=\linewidth]{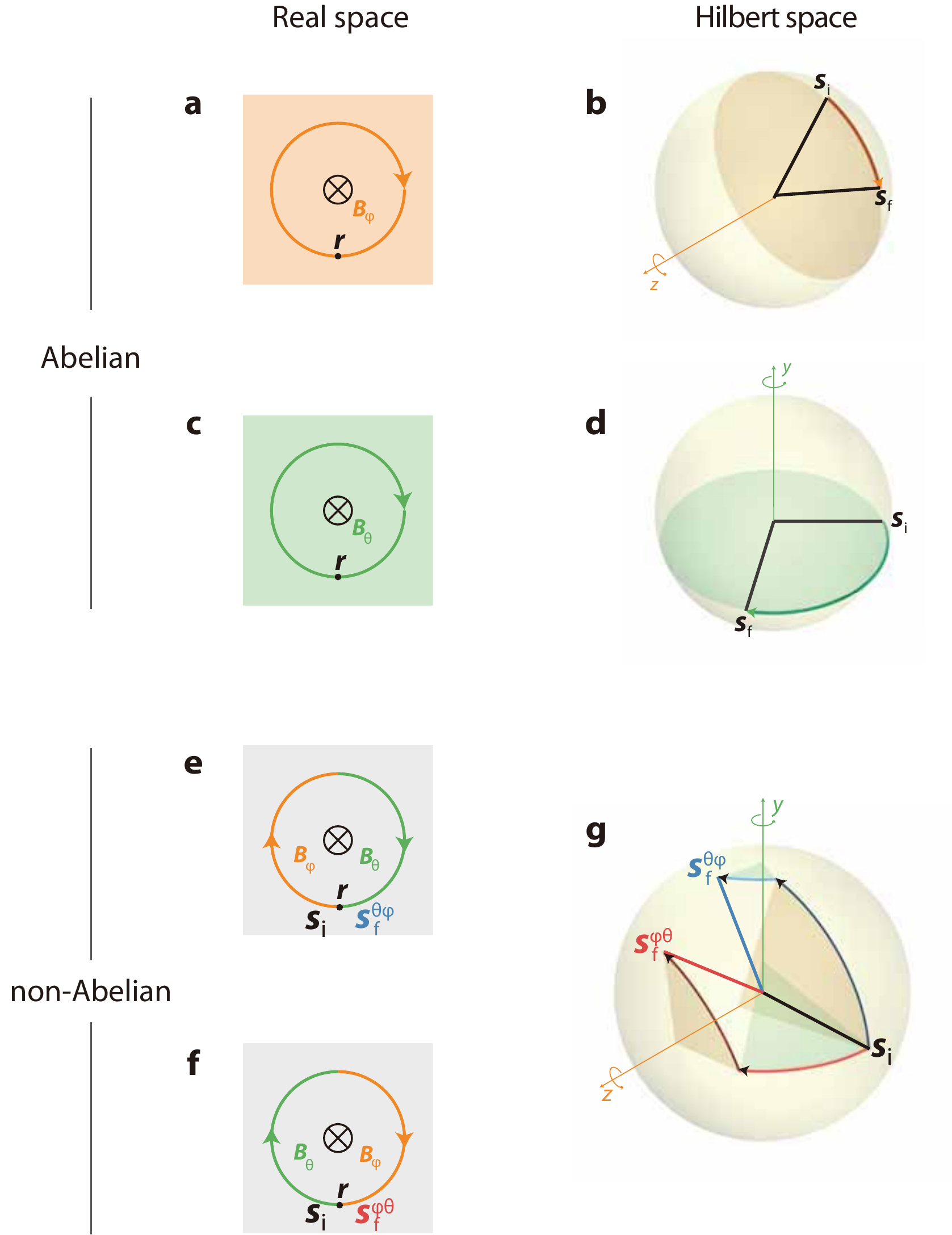}
	 \caption{%
	 	\textbf{Comparison between SU(2) Abelian and non-Abelian gauge fields in real space and in Hilbert space.}
	 	\textbf{a-d.} Along a closed loop inside an Abelian gauge field $\matr{A}\propto\sigmaz$ (\textbf{a}) or $\sigmay$ (\textbf{c}), the state evolves by rotating around the $z$ (\textbf{b}) or $y$ (\textbf{d}) axis of the Poincar\pe sphere.
	 	Within each case (\textbf{a-b} or \textbf{c-d}), the state evolution are always commutative.
        \textbf{e-f.} In non-Abelian gauge fields, the evolution operators for different loops are no longer commutative, which leads to different final states, $\matr{s}_{\text{f}}^{\theta\phi}$ and $\matr{s}_{\text{f}}^{\phi\theta}$ , for the same initial state $\matr{s}_{\text{i}}$. The non-commutativity can be tested by an Aharonov--Bohm interference of the two final states.
	    }
	\label{fig1}
\end{figure}

The difference between how a state evolves in Abelian gauge fields versus in non-Abelian gauge fields is shown in Fig.~\ref{fig1}.
In a uniform Abelian gauge field $\propto\sigmaz$, the evolution operator along a closed loop can be simplified as $\matr{W} = \e^{\iu\phi\matr{\sigma}_z}$, where $\sigmaz$ is the $z$ component of the Pauli matrices and $\phi$ is the flux of the gauge field through this closed loop (Fig.~\ref{fig1}a).
Consequently, the state rotates by $2\phi$ around the $z$ axis of the Poincar\pe sphere (Fig.~\ref{fig1}b).
If the state evolves along two consecutive closed loops, the two evolution operators are commutative, which reflects the Abelian nature of this gauge field.
Similarly, a homogeneous gauge field $\propto\sigmay$ in real space (Fig.~\ref{fig1}c) is also Abelian, as the state always evolve around the $y$ axis in the Hilbert space (Fig.~\ref{fig1}d).

In contrast, non-Abelian gauge fields require inhomogeneous gauge structures.
Fig.~\ref{fig1}ef illustrate such an example where two different $\sigmaz$ and $\sigmay$ gauge structures are concatenated into one compound closed loop.
The same initial state $\matr{s}_{\iu}$ can now evolve into different final states: $\matr{s}_{\text{f}}^{\theta\phi}$ or $\matr{s}_{\text{f}}^{\phi\theta}$ (Fig.~\ref{fig1}g), depending on the different ordering---$\phi$ and then $\theta$, or alternatively, $\theta$ and then $\phi$---of the two gauge structures.
The interference between the two final states $\matr{s}_{\text{f}}^{\theta\phi}$ and $\matr{s}_{\text{f}}^{\phi\theta}$ is known as the non-Abelian Aharonov--Bohm effect~\cite{wu1975concept,horvathy1986non,alford1990discrete,chaichian2002aharonov,iadecola2016non,chen2018non,dalibard2011colloquium,bohm2013geometric,jacob2007cold,osterloh2005cold}. This effect, that we will experimentally demonstrate later, is the most direct manifestation of non-Abelian gauge fields in real space.

\begin{figure}[htbp]
	\centering
	\includegraphics[width=1\linewidth]{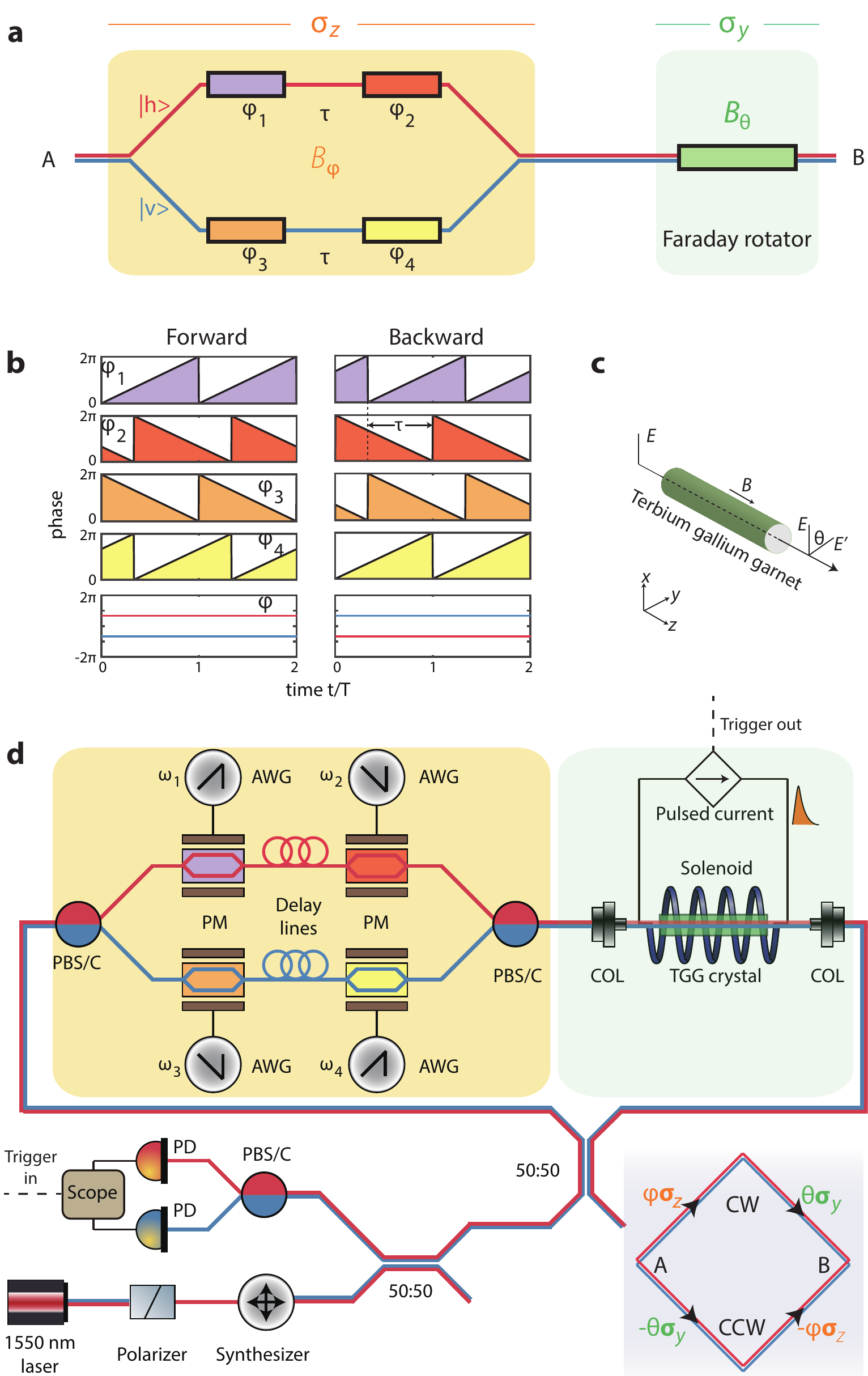}
	 \caption{%
	 	\textbf{Synthesis of non-Abelian gauge fields.}
	 	\textbf{a.} Non-Abelian gauge fields for photons. Temporal modulation and the Faraday effect, which break T-symmetry in two orthogonal bases of the Hilbert space, are used to synthesize $\sigmaz$ and $\sigmay$ gauge fields, respectively.
        \textbf{b.} Pseudospin-dependent non-reciprocal phase shifts are created through sawtooth phase modulations, which corresponds a synthetic gauge field along $\sigmaz$.
        \textbf{c.} Non-reciprocal rotation of the pseudospin is achieved via the Faraday effect in a terbium gallium garnet crystal, which corresponds to a synthetic gauge field along $\sigmay$.
	 	\textbf{d.} Experimental setup. The interference between different final pseudospin states---originated from reversed ordering of the gauge structures (CW and CCW, inset)---is read out through a Sagnac interferometer, which gives rise to the non-Abelian Aharonov---Bohm effect.
	 	PBS/C: polarization beam splitter/combiner; PM: phase modulator; AWG: arbitrary waveform generator; COL: collimator; TGG: Terbium Gallium Garnet; PD: photodetector.
	    }
	\label{fig2}
\end{figure}

In our photonic implementation, we experimentally synthesize the inhomogeneous gauge potentials in a fiber-optic system, which is conceptually illustrated in Fig.~\ref{fig2}a.
We identify the horizontal and vertical transverse modes (denoted by $\hspin$ and $\vspin$ respectively) in optical fibers as the pseudospin.
Crucially, we synthesize two types of gauge fields, $\phi\sigmaz$ and $\theta\sigmay$, using two distinct methods to break \T-symmetry.

\begin{figure*}[htbp]
	\centering
	\includegraphics[width=1\linewidth]{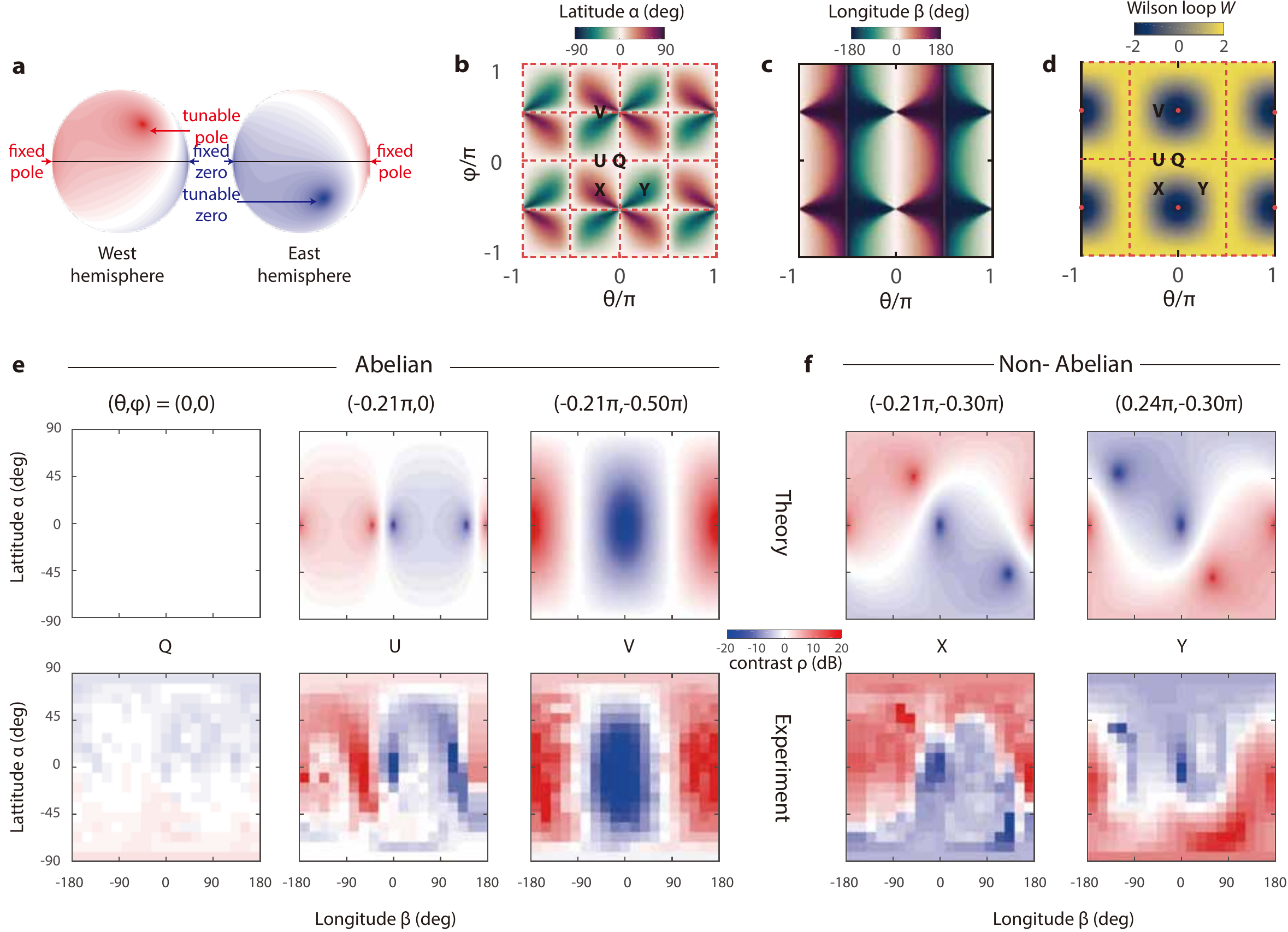}
	 \caption{%
	 	\textbf{Non-Abelian Aharonov--Bohm interference.}
	 	\textbf{a.}~Contrast function $\rho$ on the Poincar\pe sphere, featured by a fixed zero/pole pair on the equator, and a tunable zero/pole pair (which indicates the consequence of gauge fluxes).
	 	The two pairs of zeros and poles are always antipodal.
	 	\textbf{b-c.} Location (latitude and longitude) of the tunable pole on the Poincar\pe sphere as a function of the gauge fluxes $(\theta,\phi)$.
	 	Abelian gauge fields correspond to on-equator poles (red dashed lines); non-Abelian gauge fields correspond to off-equator poles---both of which are experimentally demonstrated.
        \textbf{d.}~ Wilson loops $W$ on the synthetic torus $(\theta,\phi)$. $|W|=2$ (red dashed lines) is a necessary but insufficient condition for non-Abelian guage fields (\cf \textbf{b}) .
	 	\textbf{e-f.}~Examples of predicted and observed contrast functions $\rho$ for Abelian (Q, U, and V) and non-Abelian (X and Y) gauge fields.
	 	}
	\label{fig3}
\end{figure*}

To construct a gauge field of $\phi\sigmaz$, we first employ dynamic modulations that dress $\hspin$  and $\vspin$ with nonreciprocal phase shifts of $\pm\phi$, respectively.
Specifically, four LiNbO$_3$ phase modulators---two (labeled 1 and 2) for $\hspin$ and two (labeled 3 and 4) for $\vspin$---are driven by arbitrary waveform generators that create phase shifts in the form of sawtooth functions in time (Fig. 2b).
Modulators 1 and 4 are positive in slope: $\phi_{1,4} = \omega t \mod 2\pi$; and modulators 2 and 3 are negative in slope, $\phi_{2,3} = -\omega t \mod 2\pi$.
The delay line between modulators 1 and 2 (3 and 4) corresponds to a delay time $\tau$.
As a result, besides dynamic phases, $\hspin$ ($\vspin$) picks up an extra phase $\phi=\omega\tau$ ($-\omega\tau$) in the forward (\ie left-to-right) direction, but an opposite phase $-\phi$ ($+\phi$) in the backward direction.
This pair of opposite nonreciprocal phases for opposite pseudospin components ($\hspin$ and $\vspin$) correspond to a $\phi\sigmaz$ gauge field, which is continuously tunable by varying the modulation frequency $\omega$.

A second, orthogonal type of gauge field, $\theta\sigmay$, is created using the Faraday effect.
Specifically, light is coupled out of the fiber, sent through a Terbium Gallium Garnet crystal placed in an external magnetic field, and then coupled back into the fiber.
Through the Faraday effect, pseudospin of light is rotated in a nonreciprocal way, which corresponds to a gauge field of $\theta\sigmay$. This gauge field is also continuously tunable through the external magnetic field.

We then concatenate the two non-Abelian gauge fields to demonstrate the non-Abelian Aharonov--Bohm effect via Sagnac interferometry (inset of Fig.~\ref{fig2}d).
In such Sagnac configuration, the two sites A and B in Fig.~\ref{fig2}a are combined into the same physical location to enable well-defined non-Abelian gauge fluxes.
Evolved from the clockwise (CW) and counter-clockwise (CCW) paths of the Sagnac loop, the two final states are  $\matr{s}_{\text{f}}^{\theta\phi}=\sigmaz\e^{\iu\theta\sigmay}\e^{\iu\phi\sigmaz}\matr{s}_{\iu}$ and $\matr{s}_{\text{f}}^{\phi\theta}=\e^{-\iu\phi\sigmaz}\e^{-\iu\theta\sigmay}\sigmaz\matr{s}_{\iu}$, where the $\sigmaz$ term maintains a consistent handedness of the polarization for counter-propagating states.
The interference of the two final states is given by (Sec.~S6)
\begin{align}
    \mathbf{s}_{\text{f}}=\matr{s}_{\text{f}}^{\theta\phi}+\matr{s}_{\text{f}}^{\phi\theta}
    = -\sigmax\left(\e^{\iu\theta'\sigmay}\e^{\iu\phi\sigmaz} + \e^{\iu\phi\sigmaz}\e^{\iu\theta^\prime\sigmay}\right)\matr{s}_{\iu},
\end{align}
where $\theta^\prime=\theta + \pi/2$ and $\sigmax$ is a global spin flip.
This interference describes a Sagnac-type realization of the non-Abelian Aharonov--Bohm effect~\cite{dalibard2011colloquium}---the interference between two final states, which originate from the same initial state, but undergo reversely-ordered, inhomogeneous path integrals (Fig.~\ref{fig1}e-g) in the CW and CCW directions.

Fig.~\ref{fig2}d details our experimental setup (Sec.~S1).
We place a polarization synthesizer in front of the Sagnac loop, to prepare any desired pseudospin state as the input in a deterministic manner.
After exiting the Sagnac loop, the two final states $\matr{s}_{\text{f}}^{\theta\phi}$ and $\matr{s}_{\text{f}}^{\phi\theta}$ interfere with each other.
The associated interference intensity is projected onto the horizontal and vertical bases, which are then measured separately.
Within the Sagnac loop, a solenoid---driven by tunable pulsed currents (peak current $\approx$~\SI{2}{\kA}, duration $\approx$~\SI{10}{\ms})---provides a magnetic field between 0 and $\approx$~\SI{2}{T} (Sec.~S2) for the Faraday rotator.
The solenoid also provides a temporal trigger signal for the detection.
For the dynamic modulation, we assign four different modulation frequencies (\ie slopes of the temporal sawtooth functions) $+\omega_1$, $-\omega_2$, $-\omega_3$, and $+\omega_4$ to each of the modulators with $\omega_i$ defined to be positive.
We impose an additional constraint that $\omega\equiv(\omega_1+\omega_2)/2 = (\omega_4 + \omega_3)/2$.
This modified arrangement from Fig.~\ref{fig2}b maintains the same nonreciprocal phases and thus the gauge fields $\phi\sigmaz$ (Sec.~S8).
The advantage of this modification is an experimental one: it relocates the relevant interference fringes from zero to a nonzero carrier frequency $\Omega\equiv\omega_{1}-\omega_{2}+\omega_{3}-\omega_{4}$, which is less sensitive to environmental or back-scattering noises.

We next define our experimental observable and explain its relevance to non-Abelian gauge fields.
In the original U(1) Abelian Aharonov--Bohm effect, the observable is the interference intensity as a function of the Abelian magnetic flux.
In our case, analogously, for each given set of non-Abelian gauge fluxes $(\theta,\phi)$, we measure the contrast $\rho$ between the interference intensities projected onto the horizontal and vertical bases.
Specifically, we measure $\rho (\theta,\phi,\alpha,\beta) \equiv I_{\text{h}}^\Omega/I_{\text{v}}^\Omega$, where $(\alpha,\beta)$ are the latitude and longitude of the input pseudospin state on the Poincar\'{e} sphere and $I_{\text{h(v)}}^\Omega$ is the intensity of $\ket{\text{h}}$ and $\ket{\text{v}}$ component of the output pseudospin state at the carrier frequency $\Omega$, respectively.
Therefore, $\rho$ is defined on a manifold of $S^2\times T^2$, which is spanned by the Hilbert space of the input pseudospin $S^2$ and the synthetic space of the gauge fluxes ($\theta$, $\phi$) that is $T^2$.

For a fixed set of magnetic fluxes $(\theta,\phi)$, the contrast function $\rho(\alpha,\beta)$ always exhibits two pairs of first-order zeros and poles on the Poincar\pe sphere (Fig.~\ref{fig3}a; also see Sec.~S9).
Within each pair, the zero and the pole are always antipodal and thus represent orthogonal pseudospins.
One pair, being linear polarizations $(1,0)$ (zero) and $(0,1)$ (pole), is fixed on the two ends of the equator, regardless of the choice of $(\theta,\phi$).
The other orthogonal pesudospin pair, however, is tunable on the entire sphere via the synthetic gauge fluxes $(\theta,\phi)$.
These zeros and poles are conserved quantities on the Poincar\pe sphere and dictate the behavior of the contrast $\rho$ function.
Their generation, evolution, and annihilation are directly related to the transitions between the Abelian and non-Abelian regimes.
Fig.~\ref{fig3}bc show the latitude $\alpha$ and longitude $\beta$ of the tunable pole on the Poincar\pe sphere, as a function of magnetic fluxes $(\theta,\phi)$.
When $\theta=m\pi/2$ or $\phi=n\pi/2$ ($m$ and $n$ are integers), the tunable zero-pole pair appears on the equator (red dashed lines in Fig.~\ref{fig3}b).
This key feature---an on/off-equator zero/pole---can be used to straightforwardly differentiate between Abelian and non-Abelian gauge fields synthesized in our experiment (see Fig.~\ref{fig3}ef).

The necessary and sufficient condition for gauge fields to be non-Abelian is as follows. There exists two loop operators, $\matr{W}_1$ and $\matr{W}_2$, both starting and ending at the same site in space, such that they are non-commutative, \ie $\matr{W}_1\matr{W}_2 \neq \matr{W}_2\matr{W}_1$~\cite{goldman2014light}.
In an Aharonov--Bohm interference, whether Abelian or non-Abelian, $\matr{W}_1$ and $\matr{W}_2$ can be identified as a pair of time-reversal partners that share the same physical path.
We first examine $\matr{W}_1=l_\text{b}^{-1} \cdot l_\text{t}$ and $\matr{W}_2=l_\text{t}^{-1} \cdot l_\text{b}$ in the Abelian Aharonov--Bohm experiment, whose two distinct top and bottom paths are denoted by $l_\text{t}$ and $l_\text{b}$, respectively.
Under time-reversal, both momentum and vector potential flip sign, rendering $\matr{W}_1=\matr{W}_2=\e^{\iu \gamma}$ that are clearly commutative and exhibit identical, scalar Berry phases $\gamma$ (Sec.~S7.A).
In our non-Abelian Aharonov--Bohm experiment, the time-reversal pair $\matr{W}_1$ and $\matr{W}_2$ can be analogously defined by replacing $l_\text{t}$ and $l_\text{b}$ with CW and CCW paths (Fig.~\ref{fig2}d inset), which yields (Sec.~S7.A)
\begin{align}
    \matr{W}_1 &= \mathcal{P}\exp \iu \oint_{\text{CCW}^{-1}\cdot\text{CW}} \matr{A}\diff{\matr{l}}   
    =\sigmaz\e^{\iu\theta\sigmay}\e^{\iu\phi\sigmaz}\sigmaz\e^{\iu\theta\sigmay}\e^{\iu\phi\sigmaz}, \label{eq:BerryHolo1}\\
    \matr{W}_2 &= \mathcal{P}\exp \iu \oint_{\text{CW}^{-1}\cdot\text{CCW}} \!\!\!\!-\matr{A}\diff{\matr{l}}
    =\e^{\iu\phi\sigmaz}\e^{\iu\theta\sigmay}\sigmaz\e^{\iu\phi\sigmaz}\e^{\iu\theta\sigmay}\sigmaz. \label{eq:BerryHolo2}
\end{align}
The condition for $\matr{W}_1$ and $\matr{W}_2$ to be non-commutative is satisfied when $\theta\neq m\pi/2$ and $\phi\neq n\pi/2$  (Sec.~S7.A)---the same condition also guarantees the existence of a zero and a pole of the contrast function away from the equator (Fig.~\ref{fig3}b).
$\matr{W}_1$ and $\matr{W}_2$ are also connected via a unitary gauge transformation (Sec.~S7.A); therefore they always share the same Wilson loop (Sec.~S7.C)
$
    W = \mathrm{Tr}\,\matr{W}_1= \mathrm{Tr}\,\matr{W}_2 =2 - 4\cos^2\theta\sin^2\phi. \label{eq:Wilson}
$
Fig.~\ref{fig3}d shows this Wilison loop on the $T^2$ space of gauge fluxes.
Generally speaking, in an $N$-fold degenerate system, $|W|=N$ means the state evolution can be trivially understood by decoupling the system into the product of $N$ Abelian subsystems~\cite{goldman2014light}.
In our case, such trivial configurations are shown with red dashed lines [$\theta=(m+1/2)\pi$, $\phi=n\pi$, or $\theta=m\pi$ and $\phi=(n+1/2)\pi$] in Fig.~\ref{fig3}d.
Nevertheless, $|W|\neq N$ is only a necessary but insufficient condition for gauge fields to be non-Abelian~\cite{goldman2014light}, as evident from the comparison between Fig.~\ref{fig3}b and Fig.~\ref{fig3}d: some configurations with $|W|\neq N$ are still Abelian.

\begin{figure}[t]
	\centering
	\includegraphics[width=\linewidth]{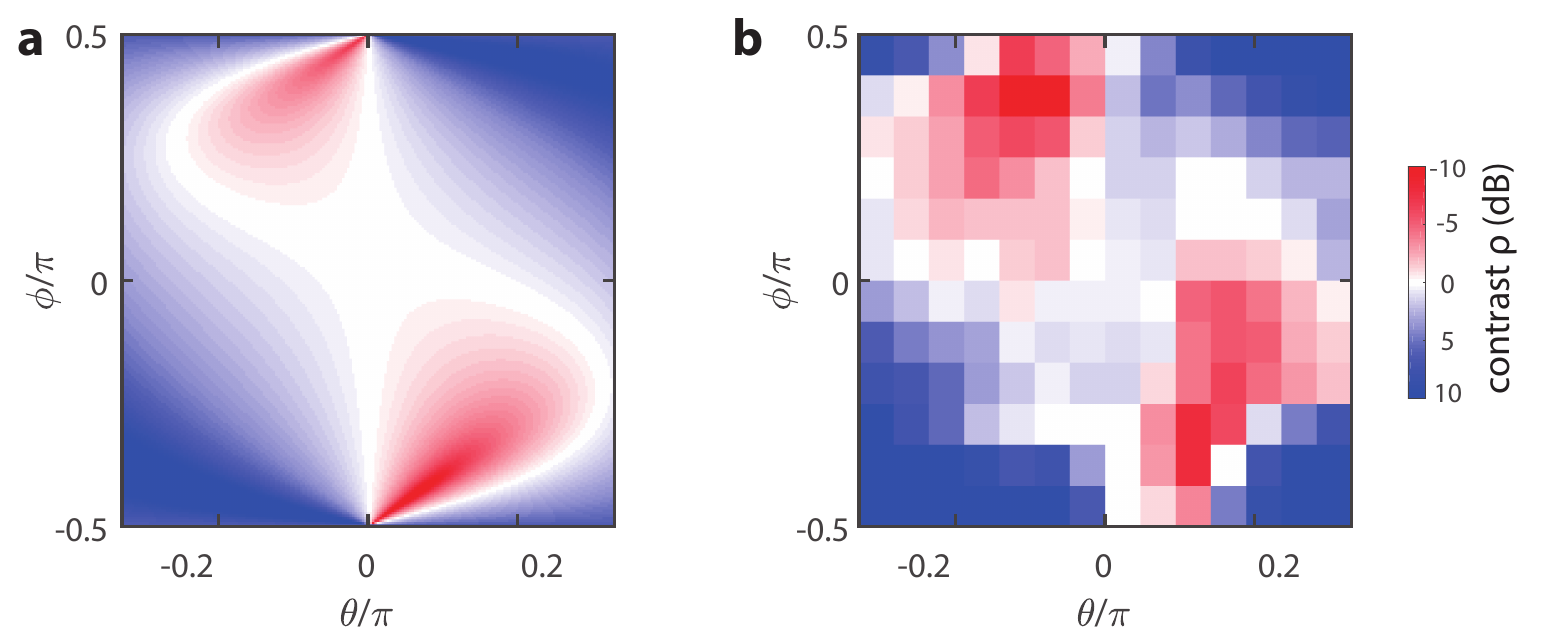}
	 \caption{%
	 	\textbf{Tunability of the non-Abelian gauge fields.}
	 	Predicted (\textbf{a}) and measured (\textbf{b}) contrast function $\rho$ for a fixed incident pseudospin state $(\alpha,\beta)\approx(-51\degree,-12\degree)$.
        The gauge fields $\phi\sigmaz$ and $\theta\sigmay$ are continuously tuned by respectively varying the modulation frequencies in the arbitrary waveform generators and the voltages applied to the solenoid.
	    }
	\label{fig4}
\end{figure}

In Fig.~\ref{fig3}ef, we characterize our synthetic gauge fields by measuring the contrast function $\rho$.
We present the comparison between theoretical predictions (top row) and experimental measurements (bottom row) for five sampling points on the synthetic space $T^2$: Q, U, V are Abelian; and X, Y are non-Abelian.
In the Abelian case Q [$(\theta,\phi)\approx(0,0)$], the tunable pole and the fixed zero annihilate each other at $(\alpha,\beta)=(0\degree,0\degree)$; so do the tunable zero and the fixed pole at $(\alpha,\beta)=(0\degree,180\degree)$.
As a result, the contrast remains a constant  $\rho = 1$ regardless of the input pseudospin state.
This is a direct consequence of the preserved \T-symmetry in the absence of gauge fluxes.
In case U [$(\theta,\phi)\approx(-0.21\pi,0)$], the annihilation of poles with zeros are lifted; nevertheless, both poles and zeros appear on the equator, and the gauge structure remains Abelian, since we only break \T-symmetry once.
In case V [$(\theta,\phi)\approx(-0.21\pi,0.50\pi)$], which is still Abelian, the two poles (zeros) coalesce and produce a second-order pole (zero) on the equator.
In cases X [$(\theta,\phi)\approx(-0.21\pi,-0.30\pi)$] and Y [$(\theta,\phi)\approx(0.24\pi,-0.30\pi)$], our synthesized gauge fields become non-Abelian, as indicated by the observed off-equator zeros and poles.
For all the cases, our observations show agreement with the associated predictions.
In our interferometer, the two spin basis $\hspin$ and $\vspin$, are not perfectly degenerate due to the difference in their refractive indices ($\sim10^{-4}$). This difference leads to a reciprocal, linear birefringent phase (\ie a dynamic phase contribution), which is calibrated and consistently applied to all measurements (Sec.~S4).

Up to this point, we have measured the contrast $\rho$ for fixed gauge fluxes, while changing the input states.
In a complementary manner, we can now fix the input state $(\alpha,\beta)$ and demonstrate the  tunability of the synthesized non-Abelian gauge fields by measuring the contrast $\rho$ for different synthetic gauge fluxes $(\theta,\phi)$. As shown in Fig.~\ref{fig4}, we reach similar agreeement between the theoretical prediction and the measurement.

In summary, we demonstrate an experimental synthesis of non-Abelian gauge fields in the real space, which is confirmed by our observation of the non-Abelian Aharonov---Bohm effect using classical particles and classical fluxes.
The realized gauge fields demonstrate a viable way to engineer the Peierls phase in the  simulation of topological systems, such as the non-Abelian Hofstadter models~\cite{osterloh2005cold,goldman2009ultracold} (also see Sec.~S10).
Our experiment also introduces non-Abelian ingredients for realizing high-order topological phases~\cite{benalcazar2017quantized,peterson2018quantized,serra2018observation} and topological pumps~\cite{zilberberg2018photonic,lohse2018exploring}.
Besides, recent advances in on-chip modulation~\cite{wang2018integrated} and magneto-optical materials~\cite{bi2018materials} could enable future observations of non-Abelian topology in  integrated photonic platforms.
Towards the quantum regime, non-Abelian gauge fields may be utilized to help generate  non-Abelian anyonic excitation~\cite{burrello2010non,keilmann2011statistically,yuan2017creating} to offer an alternative, synthetic approach for topological quantum computation.
Finally, the synergy of non-Abelian gauge fields with engineered interactions
(\eg bosonic blockade and superconducting qubits) may enable the realization of many-body physics such as the non-Abelian fractional quantum hall effect.




\bibliographystyle{apsrev4-1-fixspace} 

\begin{thebibliography}{72}%
\makeatletter
\providecommand \@ifxundefined [1]{%
 \@ifx{#1\undefined}
}%
\providecommand \@ifnum [1]{%
 \ifnum #1\expandafter \@firstoftwo
 \else \expandafter \@secondoftwo
 \fi
}%
\providecommand \@ifx [1]{%
 \ifx #1\expandafter \@firstoftwo
 \else \expandafter \@secondoftwo
 \fi
}%
\providecommand \natexlab [1]{#1}%
\providecommand \enquote  [1]{``#1''}%
\providecommand \bibnamefont  [1]{#1}%
\providecommand \bibfnamefont [1]{#1}%
\providecommand \citenamefont [1]{#1}%
\providecommand \href@noop [0]{\@secondoftwo}%
\providecommand \href [0]{\begingroup \@sanitize@url \@href}%
\providecommand \@href[1]{\@@startlink{#1}\@@href}%
\providecommand \@@href[1]{\endgroup#1\@@endlink}%
\providecommand \@sanitize@url [0]{\catcode `\\12\catcode `\$12\catcode
  `\&12\catcode `\#12\catcode `\^12\catcode `\_12\catcode `\%12\relax}%
\providecommand \@@startlink[1]{}%
\providecommand \@@endlink[0]{}%
\providecommand \url  [0]{\begingroup\@sanitize@url \@url }%
\providecommand \@url [1]{\endgroup\@href {#1}{\urlprefix }}%
\providecommand \urlprefix  [0]{URL }%
\providecommand \Eprint [0]{\href }%
\providecommand \doibase [0]{http://dx.doi.org/}%
\providecommand \selectlanguage [0]{\@gobble}%
\providecommand \bibinfo  [0]{\@secondoftwo}%
\providecommand \bibfield  [0]{\@secondoftwo}%
\providecommand \translation [1]{[#1]}%
\providecommand \BibitemOpen [0]{}%
\providecommand \bibitemStop [0]{}%
\providecommand \bibitemNoStop [0]{.\EOS\space}%
\providecommand \EOS [0]{\spacefactor3000\relax}%
\providecommand \BibitemShut  [1]{\csname bibitem#1\endcsname}%
\let\auto@bib@innerbib\@empty
\bibitem [{\citenamefont {Aharonov}\ and\ \citenamefont
  {Bohm}(1959)}]{aharonov1959significance}%
  \BibitemOpen
  \bibfield  {author} {\bibinfo {author} {\bibfnamefont {Y.}~\bibnamefont
  {Aharonov}}\ and\ \bibinfo {author} {\bibfnamefont {D.}~\bibnamefont
  {Bohm}},\ }\href {\doibase 10.1103/PhysRev.115.485} {\bibfield  {journal}
  {\bibinfo  {journal} {Phys. Rev.}\ }\textbf {\bibinfo {volume} {115}},\
  \bibinfo {pages} {485} (\bibinfo {year} {1959})}\BibitemShut {NoStop}%
\bibitem [{\citenamefont {Berry}(1984)}]{berry1984quantal}%
  \BibitemOpen
  \bibfield  {author} {\bibinfo {author} {\bibfnamefont {M.~V.}\ \bibnamefont
  {Berry}},\ }\href {\doibase 10.1098/rspa.1984.0023} {\bibfield  {journal}
  {\bibinfo  {journal} {Proc. R. Soc. Lond. A}\ }\textbf {\bibinfo {volume}
  {392}},\ \bibinfo {pages} {45} (\bibinfo {year} {1984})}\BibitemShut
  {NoStop}%
\bibitem [{\citenamefont {Pancharatnam}(1956)}]{pancharatnam1956generalized}%
  \BibitemOpen
  \bibfield  {author} {\bibinfo {author} {\bibfnamefont {S.}~\bibnamefont
  {Pancharatnam}},\ }in\ \href {\doibase 10.1007/BF03046095} {\emph {\bibinfo
  {booktitle} {Proceedings of the Indian Academy of Sciences-Section A}}},\
  Vol.~\bibinfo {volume} {44}\ (\bibinfo {organization} {Springer},\ \bibinfo
  {year} {1956})\ pp.\ \bibinfo {pages} {398--417}\BibitemShut {NoStop}%
\bibitem [{\citenamefont {Sounas}\ and\ \citenamefont
  {Al{\`u}}(2017)}]{sounas2017non}%
  \BibitemOpen
  \bibfield  {author} {\bibinfo {author} {\bibfnamefont {D.~L.}\ \bibnamefont
  {Sounas}}\ and\ \bibinfo {author} {\bibfnamefont {A.}~\bibnamefont
  {Al{\`u}}},\ }\href {\doibase 10.1038/s41566-017-0051-x} {\bibfield
  {journal} {\bibinfo  {journal} {Nat. Photon.}\ }\textbf {\bibinfo {volume}
  {11}},\ \bibinfo {pages} {774} (\bibinfo {year} {2017})}\BibitemShut
  {NoStop}%
\bibitem [{\citenamefont {Yuan}\ \emph {et~al.}(2018)\citenamefont {Yuan},
  \citenamefont {Lin}, \citenamefont {Xiao},\ and\ \citenamefont
  {Fan}}]{yuan2018synthetic}%
  \BibitemOpen
  \bibfield  {author} {\bibinfo {author} {\bibfnamefont {L.}~\bibnamefont
  {Yuan}}, \bibinfo {author} {\bibfnamefont {Q.}~\bibnamefont {Lin}}, \bibinfo
  {author} {\bibfnamefont {M.}~\bibnamefont {Xiao}},\ and\ \bibinfo {author}
  {\bibfnamefont {S.}~\bibnamefont {Fan}},\ }\href {\doibase
  10.1364/OPTICA.5.001396} {\bibfield  {journal} {\bibinfo  {journal} {Optica}\
  }\textbf {\bibinfo {volume} {5}},\ \bibinfo {pages} {1396} (\bibinfo {year}
  {2018})}\BibitemShut {NoStop}%
\bibitem [{\citenamefont {Dalibard}\ \emph {et~al.}(2011)\citenamefont
  {Dalibard}, \citenamefont {Gerbier}, \citenamefont {Juzeli{\=u}nas},\ and\
  \citenamefont {{\"O}hberg}}]{dalibard2011colloquium}%
  \BibitemOpen
  \bibfield  {author} {\bibinfo {author} {\bibfnamefont {J.}~\bibnamefont
  {Dalibard}}, \bibinfo {author} {\bibfnamefont {F.}~\bibnamefont {Gerbier}},
  \bibinfo {author} {\bibfnamefont {G.}~\bibnamefont {Juzeli{\=u}nas}},\ and\
  \bibinfo {author} {\bibfnamefont {P.}~\bibnamefont {{\"O}hberg}},\ }\href
  {\doibase 10.1103/RevModPhys.83.1523} {\bibfield  {journal} {\bibinfo
  {journal} {Rev. Mod. Phys.}\ }\textbf {\bibinfo {volume} {83}},\ \bibinfo
  {pages} {1523} (\bibinfo {year} {2011})}\BibitemShut {NoStop}%
\bibitem [{\citenamefont {Eckardt}(2017)}]{eckardt2017colloquium}%
  \BibitemOpen
  \bibfield  {author} {\bibinfo {author} {\bibfnamefont {A.}~\bibnamefont
  {Eckardt}},\ }\href {\doibase 10.1103/RevModPhys.89.011004} {\bibfield
  {journal} {\bibinfo  {journal} {Rev. Mod. Phys.}\ }\textbf {\bibinfo {volume}
  {89}},\ \bibinfo {pages} {011004} (\bibinfo {year} {2017})}\BibitemShut
  {NoStop}%
\bibitem [{\citenamefont {Goldman}\ \emph {et~al.}(2014)\citenamefont
  {Goldman}, \citenamefont {Juzeli{\=u}nas}, \citenamefont {{\"O}hberg},\ and\
  \citenamefont {Spielman}}]{goldman2014light}%
  \BibitemOpen
  \bibfield  {author} {\bibinfo {author} {\bibfnamefont {N.}~\bibnamefont
  {Goldman}}, \bibinfo {author} {\bibfnamefont {G.}~\bibnamefont
  {Juzeli{\=u}nas}}, \bibinfo {author} {\bibfnamefont {P.}~\bibnamefont
  {{\"O}hberg}},\ and\ \bibinfo {author} {\bibfnamefont {I.~B.}\ \bibnamefont
  {Spielman}},\ }\href {\doibase 10.1088/0034-4885/77/12/126401} {\bibfield
  {journal} {\bibinfo  {journal} {Reports on Progress in Physics}\ }\textbf
  {\bibinfo {volume} {77}},\ \bibinfo {pages} {126401} (\bibinfo {year}
  {2014})}\BibitemShut {NoStop}%
\bibitem [{\citenamefont {Lu}\ \emph {et~al.}(2014)\citenamefont {Lu},
  \citenamefont {Joannopoulos},\ and\ \citenamefont
  {Solja{\v{c}}i{\'c}}}]{lu2014topological}%
  \BibitemOpen
  \bibfield  {author} {\bibinfo {author} {\bibfnamefont {L.}~\bibnamefont
  {Lu}}, \bibinfo {author} {\bibfnamefont {J.~D.}\ \bibnamefont
  {Joannopoulos}},\ and\ \bibinfo {author} {\bibfnamefont {M.}~\bibnamefont
  {Solja{\v{c}}i{\'c}}},\ }\href {\doibase 10.1038/nphoton.2014.248} {\bibfield
   {journal} {\bibinfo  {journal} {Nat. Photon.}\ }\textbf {\bibinfo {volume}
  {8}},\ \bibinfo {pages} {821} (\bibinfo {year} {2014})}\BibitemShut {NoStop}%
\bibitem [{\citenamefont {Ozawa}\ \emph {et~al.}(2018)\citenamefont {Ozawa},
  \citenamefont {Price}, \citenamefont {Amo}, \citenamefont {Goldman},
  \citenamefont {Hafezi}, \citenamefont {Lu}, \citenamefont {Rechtsman},
  \citenamefont {Schuster}, \citenamefont {Simon}, \citenamefont {Zilberberg}
  \emph {et~al.}}]{ozawa2018topological}%
  \BibitemOpen
  \bibfield  {author} {\bibinfo {author} {\bibfnamefont {T.}~\bibnamefont
  {Ozawa}}, \bibinfo {author} {\bibfnamefont {H.~M.}\ \bibnamefont {Price}},
  \bibinfo {author} {\bibfnamefont {A.}~\bibnamefont {Amo}}, \bibinfo {author}
  {\bibfnamefont {N.}~\bibnamefont {Goldman}}, \bibinfo {author} {\bibfnamefont
  {M.}~\bibnamefont {Hafezi}}, \bibinfo {author} {\bibfnamefont
  {L.}~\bibnamefont {Lu}}, \bibinfo {author} {\bibfnamefont {M.}~\bibnamefont
  {Rechtsman}}, \bibinfo {author} {\bibfnamefont {D.}~\bibnamefont {Schuster}},
  \bibinfo {author} {\bibfnamefont {J.}~\bibnamefont {Simon}}, \bibinfo
  {author} {\bibfnamefont {O.}~\bibnamefont {Zilberberg}}, \emph {et~al.},\
  }\href@noop {} {\bibfield  {journal} {\bibinfo  {journal} {arXiv preprint
  arXiv:1802.04173}\ } (\bibinfo {year} {2018})}\BibitemShut {NoStop}%
\bibitem [{\citenamefont {Aidelsburger}\ \emph {et~al.}(2018)\citenamefont
  {Aidelsburger}, \citenamefont {Nascimbene},\ and\ \citenamefont
  {Goldman}}]{aidelsburger2018artificial}%
  \BibitemOpen
  \bibfield  {author} {\bibinfo {author} {\bibfnamefont {M.}~\bibnamefont
  {Aidelsburger}}, \bibinfo {author} {\bibfnamefont {S.}~\bibnamefont
  {Nascimbene}},\ and\ \bibinfo {author} {\bibfnamefont {N.}~\bibnamefont
  {Goldman}},\ }\href {\doibase 10.1016/j.crhy.2018.03.002} {\bibfield
  {journal} {\bibinfo  {journal} {Comptes Rendus Physique}\ }\textbf {\bibinfo
  {volume} {19}},\ \bibinfo {pages} {394} (\bibinfo {year} {2018})}\BibitemShut
  {NoStop}%
\bibitem [{\citenamefont {Bloch}\ \emph {et~al.}(2012)\citenamefont {Bloch},
  \citenamefont {Dalibard},\ and\ \citenamefont
  {Nascimbene}}]{bloch2012quantum}%
  \BibitemOpen
  \bibfield  {author} {\bibinfo {author} {\bibfnamefont {I.}~\bibnamefont
  {Bloch}}, \bibinfo {author} {\bibfnamefont {J.}~\bibnamefont {Dalibard}},\
  and\ \bibinfo {author} {\bibfnamefont {S.}~\bibnamefont {Nascimbene}},\
  }\href {\doibase 10.1038/nphys2259} {\bibfield  {journal} {\bibinfo
  {journal} {Nat. Phys.}\ }\textbf {\bibinfo {volume} {8}},\ \bibinfo {pages}
  {267} (\bibinfo {year} {2012})}\BibitemShut {NoStop}%
\bibitem [{\citenamefont {Aspuru-Guzik}\ and\ \citenamefont
  {Walther}(2012)}]{aspuru2012photonic}%
  \BibitemOpen
  \bibfield  {author} {\bibinfo {author} {\bibfnamefont {A.}~\bibnamefont
  {Aspuru-Guzik}}\ and\ \bibinfo {author} {\bibfnamefont {P.}~\bibnamefont
  {Walther}},\ }\href {\doibase 10.1038/nphys2253} {\bibfield  {journal}
  {\bibinfo  {journal} {Nat. Phys.}\ }\textbf {\bibinfo {volume} {8}},\
  \bibinfo {pages} {285} (\bibinfo {year} {2012})}\BibitemShut {NoStop}%
\bibitem [{\citenamefont {Tzuang}\ \emph {et~al.}(2014)\citenamefont {Tzuang},
  \citenamefont {Fang}, \citenamefont {Nussenzveig}, \citenamefont {Fan},\ and\
  \citenamefont {Lipson}}]{tzuang2014non}%
  \BibitemOpen
  \bibfield  {author} {\bibinfo {author} {\bibfnamefont {L.~D.}\ \bibnamefont
  {Tzuang}}, \bibinfo {author} {\bibfnamefont {K.}~\bibnamefont {Fang}},
  \bibinfo {author} {\bibfnamefont {P.}~\bibnamefont {Nussenzveig}}, \bibinfo
  {author} {\bibfnamefont {S.}~\bibnamefont {Fan}},\ and\ \bibinfo {author}
  {\bibfnamefont {M.}~\bibnamefont {Lipson}},\ }\href {\doibase
  10.1038/nphoton.2014.177} {\bibfield  {journal} {\bibinfo  {journal} {Nat.
  Photon.}\ }\textbf {\bibinfo {volume} {8}},\ \bibinfo {pages} {701} (\bibinfo
  {year} {2014})}\BibitemShut {NoStop}%
\bibitem [{\citenamefont {Fang}\ \emph {et~al.}(2017)\citenamefont {Fang},
  \citenamefont {Luo}, \citenamefont {Metelmann}, \citenamefont {Matheny},
  \citenamefont {Marquardt}, \citenamefont {Clerk},\ and\ \citenamefont
  {Painter}}]{fang2017generalized}%
  \BibitemOpen
  \bibfield  {author} {\bibinfo {author} {\bibfnamefont {K.}~\bibnamefont
  {Fang}}, \bibinfo {author} {\bibfnamefont {J.}~\bibnamefont {Luo}}, \bibinfo
  {author} {\bibfnamefont {A.}~\bibnamefont {Metelmann}}, \bibinfo {author}
  {\bibfnamefont {M.~H.}\ \bibnamefont {Matheny}}, \bibinfo {author}
  {\bibfnamefont {F.}~\bibnamefont {Marquardt}}, \bibinfo {author}
  {\bibfnamefont {A.~A.}\ \bibnamefont {Clerk}},\ and\ \bibinfo {author}
  {\bibfnamefont {O.}~\bibnamefont {Painter}},\ }\href {\doibase
  10.1038/nphys4009} {\bibfield  {journal} {\bibinfo  {journal} {Nat. Phys.}\
  }\textbf {\bibinfo {volume} {13}},\ \bibinfo {pages} {465} (\bibinfo {year}
  {2017})}\BibitemShut {NoStop}%
\bibitem [{\citenamefont {Lin}\ \emph {et~al.}(2009)\citenamefont {Lin},
  \citenamefont {Compton}, \citenamefont {Jimenez-Garcia}, \citenamefont
  {Porto},\ and\ \citenamefont {Spielman}}]{lin2009synthetic}%
  \BibitemOpen
  \bibfield  {author} {\bibinfo {author} {\bibfnamefont {Y.-J.}\ \bibnamefont
  {Lin}}, \bibinfo {author} {\bibfnamefont {R.~L.}\ \bibnamefont {Compton}},
  \bibinfo {author} {\bibfnamefont {K.}~\bibnamefont {Jimenez-Garcia}},
  \bibinfo {author} {\bibfnamefont {J.~V.}\ \bibnamefont {Porto}},\ and\
  \bibinfo {author} {\bibfnamefont {I.~B.}\ \bibnamefont {Spielman}},\ }\href
  {\doibase 10.1038/nature08609} {\bibfield  {journal} {\bibinfo  {journal}
  {Nature}\ }\textbf {\bibinfo {volume} {462}},\ \bibinfo {pages} {628}
  (\bibinfo {year} {2009})}\BibitemShut {NoStop}%
\bibitem [{\citenamefont {Aidelsburger}\ \emph {et~al.}(2011)\citenamefont
  {Aidelsburger}, \citenamefont {Atala}, \citenamefont {Nascimb{\`e}ne},
  \citenamefont {Trotzky}, \citenamefont {Chen},\ and\ \citenamefont
  {Bloch}}]{aidelsburger2011experimental}%
  \BibitemOpen
  \bibfield  {author} {\bibinfo {author} {\bibfnamefont {M.}~\bibnamefont
  {Aidelsburger}}, \bibinfo {author} {\bibfnamefont {M.}~\bibnamefont {Atala}},
  \bibinfo {author} {\bibfnamefont {S.}~\bibnamefont {Nascimb{\`e}ne}},
  \bibinfo {author} {\bibfnamefont {S.}~\bibnamefont {Trotzky}}, \bibinfo
  {author} {\bibfnamefont {Y.-A.}\ \bibnamefont {Chen}},\ and\ \bibinfo
  {author} {\bibfnamefont {I.}~\bibnamefont {Bloch}},\ }\href {\doibase
  10.1103/PhysRevLett.107.255301} {\bibfield  {journal} {\bibinfo  {journal}
  {Phys. Rev. Lett.}\ }\textbf {\bibinfo {volume} {107}},\ \bibinfo {pages}
  {255301} (\bibinfo {year} {2011})}\BibitemShut {NoStop}%
\bibitem [{\citenamefont {Miyake}\ \emph {et~al.}(2013)\citenamefont {Miyake},
  \citenamefont {Siviloglou}, \citenamefont {Kennedy}, \citenamefont {Burton},\
  and\ \citenamefont {Ketterle}}]{miyake2013realizing}%
  \BibitemOpen
  \bibfield  {author} {\bibinfo {author} {\bibfnamefont {H.}~\bibnamefont
  {Miyake}}, \bibinfo {author} {\bibfnamefont {G.~A.}\ \bibnamefont
  {Siviloglou}}, \bibinfo {author} {\bibfnamefont {C.~J.}\ \bibnamefont
  {Kennedy}}, \bibinfo {author} {\bibfnamefont {W.~C.}\ \bibnamefont
  {Burton}},\ and\ \bibinfo {author} {\bibfnamefont {W.}~\bibnamefont
  {Ketterle}},\ }\href {\doibase 10.1103/PhysRevLett.111.185302} {\bibfield
  {journal} {\bibinfo  {journal} {Phys. Rev. Lett.}\ }\textbf {\bibinfo
  {volume} {111}},\ \bibinfo {pages} {185302} (\bibinfo {year}
  {2013})}\BibitemShut {NoStop}%
\bibitem [{\citenamefont {Struck}\ \emph {et~al.}(2012)\citenamefont {Struck},
  \citenamefont {{\"O}lschl{\"a}ger}, \citenamefont {Weinberg}, \citenamefont
  {Hauke}, \citenamefont {Simonet}, \citenamefont {Eckardt}, \citenamefont
  {Lewenstein}, \citenamefont {Sengstock},\ and\ \citenamefont
  {Windpassinger}}]{struck2012tunable}%
  \BibitemOpen
  \bibfield  {author} {\bibinfo {author} {\bibfnamefont {J.}~\bibnamefont
  {Struck}}, \bibinfo {author} {\bibfnamefont {C.}~\bibnamefont
  {{\"O}lschl{\"a}ger}}, \bibinfo {author} {\bibfnamefont {M.}~\bibnamefont
  {Weinberg}}, \bibinfo {author} {\bibfnamefont {P.}~\bibnamefont {Hauke}},
  \bibinfo {author} {\bibfnamefont {J.}~\bibnamefont {Simonet}}, \bibinfo
  {author} {\bibfnamefont {A.}~\bibnamefont {Eckardt}}, \bibinfo {author}
  {\bibfnamefont {M.}~\bibnamefont {Lewenstein}}, \bibinfo {author}
  {\bibfnamefont {K.}~\bibnamefont {Sengstock}},\ and\ \bibinfo {author}
  {\bibfnamefont {P.}~\bibnamefont {Windpassinger}},\ }\href {\doibase
  10.1103/PhysRevLett.108.225304} {\bibfield  {journal} {\bibinfo  {journal}
  {Phys. Rev. Lett.}\ }\textbf {\bibinfo {volume} {108}},\ \bibinfo {pages}
  {225304} (\bibinfo {year} {2012})}\BibitemShut {NoStop}%
\bibitem [{\citenamefont {Aidelsburger}\ \emph {et~al.}(2013)\citenamefont
  {Aidelsburger}, \citenamefont {Atala}, \citenamefont {Lohse}, \citenamefont
  {Barreiro}, \citenamefont {Paredes},\ and\ \citenamefont
  {Bloch}}]{aidelsburger2013realization}%
  \BibitemOpen
  \bibfield  {author} {\bibinfo {author} {\bibfnamefont {M.}~\bibnamefont
  {Aidelsburger}}, \bibinfo {author} {\bibfnamefont {M.}~\bibnamefont {Atala}},
  \bibinfo {author} {\bibfnamefont {M.}~\bibnamefont {Lohse}}, \bibinfo
  {author} {\bibfnamefont {J.~T.}\ \bibnamefont {Barreiro}}, \bibinfo {author}
  {\bibfnamefont {B.}~\bibnamefont {Paredes}},\ and\ \bibinfo {author}
  {\bibfnamefont {I.}~\bibnamefont {Bloch}},\ }\href {\doibase
  10.1103/PhysRevLett.111.185301} {\bibfield  {journal} {\bibinfo  {journal}
  {Phys. Rev. Lett.}\ }\textbf {\bibinfo {volume} {111}},\ \bibinfo {pages}
  {185301} (\bibinfo {year} {2013})}\BibitemShut {NoStop}%
\bibitem [{\citenamefont {Aidelsburger}\ \emph {et~al.}(2015)\citenamefont
  {Aidelsburger}, \citenamefont {Lohse}, \citenamefont {Schweizer},
  \citenamefont {Atala}, \citenamefont {Barreiro}, \citenamefont {Nascimbene},
  \citenamefont {Cooper}, \citenamefont {Bloch},\ and\ \citenamefont
  {Goldman}}]{aidelsburger2015measuring}%
  \BibitemOpen
  \bibfield  {author} {\bibinfo {author} {\bibfnamefont {M.}~\bibnamefont
  {Aidelsburger}}, \bibinfo {author} {\bibfnamefont {M.}~\bibnamefont {Lohse}},
  \bibinfo {author} {\bibfnamefont {C.}~\bibnamefont {Schweizer}}, \bibinfo
  {author} {\bibfnamefont {M.}~\bibnamefont {Atala}}, \bibinfo {author}
  {\bibfnamefont {J.~T.}\ \bibnamefont {Barreiro}}, \bibinfo {author}
  {\bibfnamefont {S.}~\bibnamefont {Nascimbene}}, \bibinfo {author}
  {\bibfnamefont {N.}~\bibnamefont {Cooper}}, \bibinfo {author} {\bibfnamefont
  {I.}~\bibnamefont {Bloch}},\ and\ \bibinfo {author} {\bibfnamefont
  {N.}~\bibnamefont {Goldman}},\ }\href {\doibase 10.1038/nphys3171} {\bibfield
   {journal} {\bibinfo  {journal} {Nat. Phys.}\ }\textbf {\bibinfo {volume}
  {11}},\ \bibinfo {pages} {162} (\bibinfo {year} {2015})}\BibitemShut
  {NoStop}%
\bibitem [{\citenamefont {Jotzu}\ \emph {et~al.}(2014)\citenamefont {Jotzu},
  \citenamefont {Messer}, \citenamefont {Desbuquois}, \citenamefont {Lebrat},
  \citenamefont {Uehlinger}, \citenamefont {Greif},\ and\ \citenamefont
  {Esslinger}}]{jotzu2014experimental}%
  \BibitemOpen
  \bibfield  {author} {\bibinfo {author} {\bibfnamefont {G.}~\bibnamefont
  {Jotzu}}, \bibinfo {author} {\bibfnamefont {M.}~\bibnamefont {Messer}},
  \bibinfo {author} {\bibfnamefont {R.}~\bibnamefont {Desbuquois}}, \bibinfo
  {author} {\bibfnamefont {M.}~\bibnamefont {Lebrat}}, \bibinfo {author}
  {\bibfnamefont {T.}~\bibnamefont {Uehlinger}}, \bibinfo {author}
  {\bibfnamefont {D.}~\bibnamefont {Greif}},\ and\ \bibinfo {author}
  {\bibfnamefont {T.}~\bibnamefont {Esslinger}},\ }\href {\doibase
  10.1038/nature13915} {\bibfield  {journal} {\bibinfo  {journal} {Nature}\
  }\textbf {\bibinfo {volume} {515}},\ \bibinfo {pages} {237} (\bibinfo {year}
  {2014})}\BibitemShut {NoStop}%
\bibitem [{\citenamefont {Ray}\ \emph {et~al.}(2014)\citenamefont {Ray},
  \citenamefont {Ruokokoski}, \citenamefont {Kandel}, \citenamefont
  {M{\"o}tt{\"o}nen},\ and\ \citenamefont {Hall}}]{ray2014observation}%
  \BibitemOpen
  \bibfield  {author} {\bibinfo {author} {\bibfnamefont {M.~W.}\ \bibnamefont
  {Ray}}, \bibinfo {author} {\bibfnamefont {E.}~\bibnamefont {Ruokokoski}},
  \bibinfo {author} {\bibfnamefont {S.}~\bibnamefont {Kandel}}, \bibinfo
  {author} {\bibfnamefont {M.}~\bibnamefont {M{\"o}tt{\"o}nen}},\ and\ \bibinfo
  {author} {\bibfnamefont {D.}~\bibnamefont {Hall}},\ }\href {\doibase
  10.1038/nature12954} {\bibfield  {journal} {\bibinfo  {journal} {Nature}\
  }\textbf {\bibinfo {volume} {505}},\ \bibinfo {pages} {657} (\bibinfo {year}
  {2014})}\BibitemShut {NoStop}%
\bibitem [{\citenamefont {Li}\ \emph {et~al.}(2016)\citenamefont {Li},
  \citenamefont {Duca}, \citenamefont {Reitter}, \citenamefont {Grusdt},
  \citenamefont {Demler}, \citenamefont {Endres}, \citenamefont
  {Schleier-Smith}, \citenamefont {Bloch},\ and\ \citenamefont
  {Schneider}}]{li2016bloch}%
  \BibitemOpen
  \bibfield  {author} {\bibinfo {author} {\bibfnamefont {T.}~\bibnamefont
  {Li}}, \bibinfo {author} {\bibfnamefont {L.}~\bibnamefont {Duca}}, \bibinfo
  {author} {\bibfnamefont {M.}~\bibnamefont {Reitter}}, \bibinfo {author}
  {\bibfnamefont {F.}~\bibnamefont {Grusdt}}, \bibinfo {author} {\bibfnamefont
  {E.}~\bibnamefont {Demler}}, \bibinfo {author} {\bibfnamefont
  {M.}~\bibnamefont {Endres}}, \bibinfo {author} {\bibfnamefont
  {M.}~\bibnamefont {Schleier-Smith}}, \bibinfo {author} {\bibfnamefont
  {I.}~\bibnamefont {Bloch}},\ and\ \bibinfo {author} {\bibfnamefont
  {U.}~\bibnamefont {Schneider}},\ }\href {\doibase 10.1126/science.aad5812}
  {\bibfield  {journal} {\bibinfo  {journal} {Science}\ }\textbf {\bibinfo
  {volume} {352}},\ \bibinfo {pages} {1094} (\bibinfo {year}
  {2016})}\BibitemShut {NoStop}%
\bibitem [{\citenamefont {Fang}\ \emph
  {et~al.}(2012{\natexlab{a}})\citenamefont {Fang}, \citenamefont {Yu},\ and\
  \citenamefont {Fan}}]{fang2012realizing}%
  \BibitemOpen
  \bibfield  {author} {\bibinfo {author} {\bibfnamefont {K.}~\bibnamefont
  {Fang}}, \bibinfo {author} {\bibfnamefont {Z.}~\bibnamefont {Yu}},\ and\
  \bibinfo {author} {\bibfnamefont {S.}~\bibnamefont {Fan}},\ }\href {\doibase
  10.1038/nphoton.2012.236} {\bibfield  {journal} {\bibinfo  {journal} {Nat.
  Photon.}\ }\textbf {\bibinfo {volume} {6}},\ \bibinfo {pages} {782} (\bibinfo
  {year} {2012}{\natexlab{a}})}\BibitemShut {NoStop}%
\bibitem [{\citenamefont {Fang}\ \emph
  {et~al.}(2012{\natexlab{b}})\citenamefont {Fang}, \citenamefont {Yu},\ and\
  \citenamefont {Fan}}]{fang2012photonic}%
  \BibitemOpen
  \bibfield  {author} {\bibinfo {author} {\bibfnamefont {K.}~\bibnamefont
  {Fang}}, \bibinfo {author} {\bibfnamefont {Z.}~\bibnamefont {Yu}},\ and\
  \bibinfo {author} {\bibfnamefont {S.}~\bibnamefont {Fan}},\ }\href {\doibase
  10.1103/PhysRevLett.108.153901} {\bibfield  {journal} {\bibinfo  {journal}
  {Phys. Rev. Lett.}\ }\textbf {\bibinfo {volume} {108}},\ \bibinfo {pages}
  {153901} (\bibinfo {year} {2012}{\natexlab{b}})}\BibitemShut {NoStop}%
\bibitem [{\citenamefont {Hafezi}\ \emph {et~al.}(2011)\citenamefont {Hafezi},
  \citenamefont {Demler}, \citenamefont {Lukin},\ and\ \citenamefont
  {Taylor}}]{hafezi2011robust}%
  \BibitemOpen
  \bibfield  {author} {\bibinfo {author} {\bibfnamefont {M.}~\bibnamefont
  {Hafezi}}, \bibinfo {author} {\bibfnamefont {E.~A.}\ \bibnamefont {Demler}},
  \bibinfo {author} {\bibfnamefont {M.~D.}\ \bibnamefont {Lukin}},\ and\
  \bibinfo {author} {\bibfnamefont {J.~M.}\ \bibnamefont {Taylor}},\ }\href
  {\doibase 10.1038/nphys2063} {\bibfield  {journal} {\bibinfo  {journal} {Nat.
  Phys.}\ }\textbf {\bibinfo {volume} {7}},\ \bibinfo {pages} {907} (\bibinfo
  {year} {2011})}\BibitemShut {NoStop}%
\bibitem [{\citenamefont {Umucal{\i}lar}\ and\ \citenamefont
  {Carusotto}(2011)}]{umucalilar2011artificial}%
  \BibitemOpen
  \bibfield  {author} {\bibinfo {author} {\bibfnamefont {R.}~\bibnamefont
  {Umucal{\i}lar}}\ and\ \bibinfo {author} {\bibfnamefont {I.}~\bibnamefont
  {Carusotto}},\ }\href {\doibase 10.1103/PhysRevA.84.043804} {\bibfield
  {journal} {\bibinfo  {journal} {Phys. Rev. A}\ }\textbf {\bibinfo {volume}
  {84}},\ \bibinfo {pages} {043804} (\bibinfo {year} {2011})}\BibitemShut
  {NoStop}%
\bibitem [{\citenamefont {Li}\ \emph {et~al.}(2014)\citenamefont {Li},
  \citenamefont {Eggleton}, \citenamefont {Fang},\ and\ \citenamefont
  {Fan}}]{li2014photonic}%
  \BibitemOpen
  \bibfield  {author} {\bibinfo {author} {\bibfnamefont {E.}~\bibnamefont
  {Li}}, \bibinfo {author} {\bibfnamefont {B.~J.}\ \bibnamefont {Eggleton}},
  \bibinfo {author} {\bibfnamefont {K.}~\bibnamefont {Fang}},\ and\ \bibinfo
  {author} {\bibfnamefont {S.}~\bibnamefont {Fan}},\ }\href {\doibase
  10.1038/ncomms4225} {\bibfield  {journal} {\bibinfo  {journal} {Nat.
  Commun.}\ }\textbf {\bibinfo {volume} {5}},\ \bibinfo {pages} {3225}
  (\bibinfo {year} {2014})}\BibitemShut {NoStop}%
\bibitem [{\citenamefont {Mittal}\ \emph {et~al.}(2014)\citenamefont {Mittal},
  \citenamefont {Fan}, \citenamefont {Faez}, \citenamefont {Migdall},
  \citenamefont {Taylor},\ and\ \citenamefont
  {Hafezi}}]{mittal2014topologically}%
  \BibitemOpen
  \bibfield  {author} {\bibinfo {author} {\bibfnamefont {S.}~\bibnamefont
  {Mittal}}, \bibinfo {author} {\bibfnamefont {J.}~\bibnamefont {Fan}},
  \bibinfo {author} {\bibfnamefont {S.}~\bibnamefont {Faez}}, \bibinfo {author}
  {\bibfnamefont {A.}~\bibnamefont {Migdall}}, \bibinfo {author} {\bibfnamefont
  {J.}~\bibnamefont {Taylor}},\ and\ \bibinfo {author} {\bibfnamefont
  {M.}~\bibnamefont {Hafezi}},\ }\href {\doibase
  10.1103/PhysRevLett.113.087403} {\bibfield  {journal} {\bibinfo  {journal}
  {Phys. Rev. Lett.}\ }\textbf {\bibinfo {volume} {113}},\ \bibinfo {pages}
  {087403} (\bibinfo {year} {2014})}\BibitemShut {NoStop}%
\bibitem [{\citenamefont {Schine}\ \emph {et~al.}(2016)\citenamefont {Schine},
  \citenamefont {Ryou}, \citenamefont {Gromov}, \citenamefont {Sommer},\ and\
  \citenamefont {Simon}}]{schine2016synthetic}%
  \BibitemOpen
  \bibfield  {author} {\bibinfo {author} {\bibfnamefont {N.}~\bibnamefont
  {Schine}}, \bibinfo {author} {\bibfnamefont {A.}~\bibnamefont {Ryou}},
  \bibinfo {author} {\bibfnamefont {A.}~\bibnamefont {Gromov}}, \bibinfo
  {author} {\bibfnamefont {A.}~\bibnamefont {Sommer}},\ and\ \bibinfo {author}
  {\bibfnamefont {J.}~\bibnamefont {Simon}},\ }\href {\doibase
  10.1038/nature17943} {\bibfield  {journal} {\bibinfo  {journal} {Nature}\
  }\textbf {\bibinfo {volume} {534}},\ \bibinfo {pages} {671} (\bibinfo {year}
  {2016})}\BibitemShut {NoStop}%
\bibitem [{\citenamefont {Rechtsman}\ \emph {et~al.}(2013)\citenamefont
  {Rechtsman}, \citenamefont {Zeuner}, \citenamefont {T{\"u}nnermann},
  \citenamefont {Nolte}, \citenamefont {Segev},\ and\ \citenamefont
  {Szameit}}]{rechtsman2013strain}%
  \BibitemOpen
  \bibfield  {author} {\bibinfo {author} {\bibfnamefont {M.~C.}\ \bibnamefont
  {Rechtsman}}, \bibinfo {author} {\bibfnamefont {J.~M.}\ \bibnamefont
  {Zeuner}}, \bibinfo {author} {\bibfnamefont {A.}~\bibnamefont
  {T{\"u}nnermann}}, \bibinfo {author} {\bibfnamefont {S.}~\bibnamefont
  {Nolte}}, \bibinfo {author} {\bibfnamefont {M.}~\bibnamefont {Segev}},\ and\
  \bibinfo {author} {\bibfnamefont {A.}~\bibnamefont {Szameit}},\ }\href
  {\doibase 10.1038/nphoton.2012.302} {\bibfield  {journal} {\bibinfo
  {journal} {Nat. Photon.}\ }\textbf {\bibinfo {volume} {7}},\ \bibinfo {pages}
  {153} (\bibinfo {year} {2013})}\BibitemShut {NoStop}%
\bibitem [{\citenamefont {Haldane}\ and\ \citenamefont
  {Raghu}(2008)}]{haldane2008possible}%
  \BibitemOpen
  \bibfield  {author} {\bibinfo {author} {\bibfnamefont {F.}~\bibnamefont
  {Haldane}}\ and\ \bibinfo {author} {\bibfnamefont {S.}~\bibnamefont
  {Raghu}},\ }\href {\doibase 10.1103/PhysRevLett.100.013904} {\bibfield
  {journal} {\bibinfo  {journal} {Phys. Rev. Lett.}\ }\textbf {\bibinfo
  {volume} {100}},\ \bibinfo {pages} {013904} (\bibinfo {year}
  {2008})}\BibitemShut {NoStop}%
\bibitem [{\citenamefont {Wang}\ \emph {et~al.}(2009)\citenamefont {Wang},
  \citenamefont {Chong}, \citenamefont {Joannopoulos},\ and\ \citenamefont
  {Solja{\v{c}}i{\'c}}}]{wang2009observation}%
  \BibitemOpen
  \bibfield  {author} {\bibinfo {author} {\bibfnamefont {Z.}~\bibnamefont
  {Wang}}, \bibinfo {author} {\bibfnamefont {Y.}~\bibnamefont {Chong}},
  \bibinfo {author} {\bibfnamefont {J.~D.}\ \bibnamefont {Joannopoulos}},\ and\
  \bibinfo {author} {\bibfnamefont {M.}~\bibnamefont {Solja{\v{c}}i{\'c}}},\
  }\href {\doibase 10.1038/nature08293} {\bibfield  {journal} {\bibinfo
  {journal} {Nature}\ }\textbf {\bibinfo {volume} {461}},\ \bibinfo {pages}
  {772} (\bibinfo {year} {2009})}\BibitemShut {NoStop}%
\bibitem [{\citenamefont {Xiao}\ \emph {et~al.}(2015)\citenamefont {Xiao},
  \citenamefont {Chen}, \citenamefont {He},\ and\ \citenamefont
  {Chan}}]{xiao2015synthetic}%
  \BibitemOpen
  \bibfield  {author} {\bibinfo {author} {\bibfnamefont {M.}~\bibnamefont
  {Xiao}}, \bibinfo {author} {\bibfnamefont {W.-J.}\ \bibnamefont {Chen}},
  \bibinfo {author} {\bibfnamefont {W.-Y.}\ \bibnamefont {He}},\ and\ \bibinfo
  {author} {\bibfnamefont {C.~T.}\ \bibnamefont {Chan}},\ }\href {\doibase
  10.1038/nphys3458} {\bibfield  {journal} {\bibinfo  {journal} {Nat. Phys.}\
  }\textbf {\bibinfo {volume} {11}},\ \bibinfo {pages} {920} (\bibinfo {year}
  {2015})}\BibitemShut {NoStop}%
\bibitem [{\citenamefont {Yang}\ \emph {et~al.}(2017)\citenamefont {Yang},
  \citenamefont {Gao}, \citenamefont {Yang},\ and\ \citenamefont
  {Zhang}}]{yang2017strain}%
  \BibitemOpen
  \bibfield  {author} {\bibinfo {author} {\bibfnamefont {Z.}~\bibnamefont
  {Yang}}, \bibinfo {author} {\bibfnamefont {F.}~\bibnamefont {Gao}}, \bibinfo
  {author} {\bibfnamefont {Y.}~\bibnamefont {Yang}},\ and\ \bibinfo {author}
  {\bibfnamefont {B.}~\bibnamefont {Zhang}},\ }\href {\doibase
  10.1103/PhysRevLett.118.194301} {\bibfield  {journal} {\bibinfo  {journal}
  {Phys. Rev. Lett.}\ }\textbf {\bibinfo {volume} {118}},\ \bibinfo {pages}
  {194301} (\bibinfo {year} {2017})}\BibitemShut {NoStop}%
\bibitem [{\citenamefont {Abbaszadeh}\ \emph {et~al.}(2017)\citenamefont
  {Abbaszadeh}, \citenamefont {Souslov}, \citenamefont {Paulose}, \citenamefont
  {Schomerus},\ and\ \citenamefont {Vitelli}}]{abbaszadeh2017sonic}%
  \BibitemOpen
  \bibfield  {author} {\bibinfo {author} {\bibfnamefont {H.}~\bibnamefont
  {Abbaszadeh}}, \bibinfo {author} {\bibfnamefont {A.}~\bibnamefont {Souslov}},
  \bibinfo {author} {\bibfnamefont {J.}~\bibnamefont {Paulose}}, \bibinfo
  {author} {\bibfnamefont {H.}~\bibnamefont {Schomerus}},\ and\ \bibinfo
  {author} {\bibfnamefont {V.}~\bibnamefont {Vitelli}},\ }\href {\doibase
  10.1103/PhysRevLett.119.195502} {\bibfield  {journal} {\bibinfo  {journal}
  {Phys. Rev. Lett.}\ }\textbf {\bibinfo {volume} {119}},\ \bibinfo {pages}
  {195502} (\bibinfo {year} {2017})}\BibitemShut {NoStop}%
\bibitem [{\citenamefont {Lim}\ \emph {et~al.}(2017)\citenamefont {Lim},
  \citenamefont {Togan}, \citenamefont {Kroner}, \citenamefont
  {Miguel-Sanchez},\ and\ \citenamefont
  {Imamo{\u{g}}lu}}]{lim2017electrically}%
  \BibitemOpen
  \bibfield  {author} {\bibinfo {author} {\bibfnamefont {H.-T.}\ \bibnamefont
  {Lim}}, \bibinfo {author} {\bibfnamefont {E.}~\bibnamefont {Togan}}, \bibinfo
  {author} {\bibfnamefont {M.}~\bibnamefont {Kroner}}, \bibinfo {author}
  {\bibfnamefont {J.}~\bibnamefont {Miguel-Sanchez}},\ and\ \bibinfo {author}
  {\bibfnamefont {A.}~\bibnamefont {Imamo{\u{g}}lu}},\ }\href {\doibase
  10.1038/ncomms14540} {\bibfield  {journal} {\bibinfo  {journal} {Nat.
  Commun.}\ }\textbf {\bibinfo {volume} {8}},\ \bibinfo {pages} {14540}
  (\bibinfo {year} {2017})}\BibitemShut {NoStop}%
\bibitem [{\citenamefont {Schroer}\ \emph {et~al.}(2014)\citenamefont
  {Schroer}, \citenamefont {Kolodrubetz}, \citenamefont {Kindel}, \citenamefont
  {Sandberg}, \citenamefont {Gao}, \citenamefont {Vissers}, \citenamefont
  {Pappas}, \citenamefont {Polkovnikov},\ and\ \citenamefont
  {Lehnert}}]{schroer2014measuring}%
  \BibitemOpen
  \bibfield  {author} {\bibinfo {author} {\bibfnamefont {M.}~\bibnamefont
  {Schroer}}, \bibinfo {author} {\bibfnamefont {M.}~\bibnamefont
  {Kolodrubetz}}, \bibinfo {author} {\bibfnamefont {W.}~\bibnamefont {Kindel}},
  \bibinfo {author} {\bibfnamefont {M.}~\bibnamefont {Sandberg}}, \bibinfo
  {author} {\bibfnamefont {J.}~\bibnamefont {Gao}}, \bibinfo {author}
  {\bibfnamefont {M.}~\bibnamefont {Vissers}}, \bibinfo {author} {\bibfnamefont
  {D.}~\bibnamefont {Pappas}}, \bibinfo {author} {\bibfnamefont
  {A.}~\bibnamefont {Polkovnikov}},\ and\ \bibinfo {author} {\bibfnamefont
  {K.}~\bibnamefont {Lehnert}},\ }\href {\doibase
  10.1103/PhysRevLett.113.050402} {\bibfield  {journal} {\bibinfo  {journal}
  {Phys. Rev. Lett.}\ }\textbf {\bibinfo {volume} {113}},\ \bibinfo {pages}
  {050402} (\bibinfo {year} {2014})}\BibitemShut {NoStop}%
\bibitem [{\citenamefont {Roushan}\ \emph {et~al.}(2014)\citenamefont
  {Roushan}, \citenamefont {Neill}, \citenamefont {Chen}, \citenamefont
  {Kolodrubetz}, \citenamefont {Quintana}, \citenamefont {Leung}, \citenamefont
  {Fang}, \citenamefont {Barends}, \citenamefont {Campbell}, \citenamefont
  {Chen} \emph {et~al.}}]{roushan2014observation}%
  \BibitemOpen
  \bibfield  {author} {\bibinfo {author} {\bibfnamefont {P.}~\bibnamefont
  {Roushan}}, \bibinfo {author} {\bibfnamefont {C.}~\bibnamefont {Neill}},
  \bibinfo {author} {\bibfnamefont {Y.}~\bibnamefont {Chen}}, \bibinfo {author}
  {\bibfnamefont {M.}~\bibnamefont {Kolodrubetz}}, \bibinfo {author}
  {\bibfnamefont {C.}~\bibnamefont {Quintana}}, \bibinfo {author}
  {\bibfnamefont {N.}~\bibnamefont {Leung}}, \bibinfo {author} {\bibfnamefont
  {M.}~\bibnamefont {Fang}}, \bibinfo {author} {\bibfnamefont {R.}~\bibnamefont
  {Barends}}, \bibinfo {author} {\bibfnamefont {B.}~\bibnamefont {Campbell}},
  \bibinfo {author} {\bibfnamefont {Z.}~\bibnamefont {Chen}}, \emph {et~al.},\
  }\href {\doibase 10.1038/nature13891} {\bibfield  {journal} {\bibinfo
  {journal} {Nature}\ }\textbf {\bibinfo {volume} {515}},\ \bibinfo {pages}
  {241} (\bibinfo {year} {2014})}\BibitemShut {NoStop}%
\bibitem [{\citenamefont {Roushan}\ \emph {et~al.}(2017)\citenamefont
  {Roushan}, \citenamefont {Neill}, \citenamefont {Megrant}, \citenamefont
  {Chen}, \citenamefont {Babbush}, \citenamefont {Barends}, \citenamefont
  {Campbell}, \citenamefont {Chen}, \citenamefont {Chiaro}, \citenamefont
  {Dunsworth} \emph {et~al.}}]{roushan2017chiral}%
  \BibitemOpen
  \bibfield  {author} {\bibinfo {author} {\bibfnamefont {P.}~\bibnamefont
  {Roushan}}, \bibinfo {author} {\bibfnamefont {C.}~\bibnamefont {Neill}},
  \bibinfo {author} {\bibfnamefont {A.}~\bibnamefont {Megrant}}, \bibinfo
  {author} {\bibfnamefont {Y.}~\bibnamefont {Chen}}, \bibinfo {author}
  {\bibfnamefont {R.}~\bibnamefont {Babbush}}, \bibinfo {author} {\bibfnamefont
  {R.}~\bibnamefont {Barends}}, \bibinfo {author} {\bibfnamefont
  {B.}~\bibnamefont {Campbell}}, \bibinfo {author} {\bibfnamefont
  {Z.}~\bibnamefont {Chen}}, \bibinfo {author} {\bibfnamefont {B.}~\bibnamefont
  {Chiaro}}, \bibinfo {author} {\bibfnamefont {A.}~\bibnamefont {Dunsworth}},
  \emph {et~al.},\ }\href {\doibase 10.1038/nphys3930} {\bibfield  {journal}
  {\bibinfo  {journal} {Nat. Phys.}\ }\textbf {\bibinfo {volume} {13}},\
  \bibinfo {pages} {146} (\bibinfo {year} {2017})}\BibitemShut {NoStop}%
\bibitem [{\citenamefont {Huang}\ \emph {et~al.}(2016)\citenamefont {Huang},
  \citenamefont {Meng}, \citenamefont {Wang}, \citenamefont {Peng},
  \citenamefont {Zhang}, \citenamefont {Chen}, \citenamefont {Li},
  \citenamefont {Zhou},\ and\ \citenamefont {Zhang}}]{huang2016experimental}%
  \BibitemOpen
  \bibfield  {author} {\bibinfo {author} {\bibfnamefont {L.}~\bibnamefont
  {Huang}}, \bibinfo {author} {\bibfnamefont {Z.}~\bibnamefont {Meng}},
  \bibinfo {author} {\bibfnamefont {P.}~\bibnamefont {Wang}}, \bibinfo {author}
  {\bibfnamefont {P.}~\bibnamefont {Peng}}, \bibinfo {author} {\bibfnamefont
  {S.-L.}\ \bibnamefont {Zhang}}, \bibinfo {author} {\bibfnamefont
  {L.}~\bibnamefont {Chen}}, \bibinfo {author} {\bibfnamefont {D.}~\bibnamefont
  {Li}}, \bibinfo {author} {\bibfnamefont {Q.}~\bibnamefont {Zhou}},\ and\
  \bibinfo {author} {\bibfnamefont {J.}~\bibnamefont {Zhang}},\ }\href
  {\doibase 10.1038/nphys3672} {\bibfield  {journal} {\bibinfo  {journal} {Nat.
  Phys.}\ }\textbf {\bibinfo {volume} {12}},\ \bibinfo {pages} {540} (\bibinfo
  {year} {2016})}\BibitemShut {NoStop}%
\bibitem [{\citenamefont {Wu}\ \emph {et~al.}(2016)\citenamefont {Wu},
  \citenamefont {Zhang}, \citenamefont {Sun}, \citenamefont {Xu}, \citenamefont
  {Wang}, \citenamefont {Ji}, \citenamefont {Deng}, \citenamefont {Chen},
  \citenamefont {Liu},\ and\ \citenamefont {Pan}}]{wu2016realization}%
  \BibitemOpen
  \bibfield  {author} {\bibinfo {author} {\bibfnamefont {Z.}~\bibnamefont
  {Wu}}, \bibinfo {author} {\bibfnamefont {L.}~\bibnamefont {Zhang}}, \bibinfo
  {author} {\bibfnamefont {W.}~\bibnamefont {Sun}}, \bibinfo {author}
  {\bibfnamefont {X.-T.}\ \bibnamefont {Xu}}, \bibinfo {author} {\bibfnamefont
  {B.-Z.}\ \bibnamefont {Wang}}, \bibinfo {author} {\bibfnamefont {S.-C.}\
  \bibnamefont {Ji}}, \bibinfo {author} {\bibfnamefont {Y.}~\bibnamefont
  {Deng}}, \bibinfo {author} {\bibfnamefont {S.}~\bibnamefont {Chen}}, \bibinfo
  {author} {\bibfnamefont {X.-J.}\ \bibnamefont {Liu}},\ and\ \bibinfo {author}
  {\bibfnamefont {J.-W.}\ \bibnamefont {Pan}},\ }\href {\doibase
  10.1126/science.aaf6689} {\bibfield  {journal} {\bibinfo  {journal}
  {Science}\ }\textbf {\bibinfo {volume} {354}},\ \bibinfo {pages} {83}
  (\bibinfo {year} {2016})}\BibitemShut {NoStop}%
\bibitem [{\citenamefont {Wilczek}\ and\ \citenamefont
  {Zee}(1984)}]{wilczek1984appearance}%
  \BibitemOpen
  \bibfield  {author} {\bibinfo {author} {\bibfnamefont {F.}~\bibnamefont
  {Wilczek}}\ and\ \bibinfo {author} {\bibfnamefont {A.}~\bibnamefont {Zee}},\
  }\href {\doibase 10.1103/PhysRevLett.52.2111} {\bibfield  {journal} {\bibinfo
   {journal} {Phys. Rev. Lett.}\ }\textbf {\bibinfo {volume} {52}},\ \bibinfo
  {pages} {2111} (\bibinfo {year} {1984})}\BibitemShut {NoStop}%
\bibitem [{\citenamefont {Wilczek}\ and\ \citenamefont
  {Shapere}(1989)}]{wilczek1989geometric}%
  \BibitemOpen
  \bibfield  {author} {\bibinfo {author} {\bibfnamefont {F.}~\bibnamefont
  {Wilczek}}\ and\ \bibinfo {author} {\bibfnamefont {A.}~\bibnamefont
  {Shapere}},\ }\href@noop {} {\emph {\bibinfo {title} {Geometric phases in
  physics}}},\ Vol.~\bibinfo {volume} {5}\ (\bibinfo  {publisher} {World
  Scientific},\ \bibinfo {year} {1989})\BibitemShut {NoStop}%
\bibitem [{\citenamefont {Zwanziger}\ \emph {et~al.}(1990)\citenamefont
  {Zwanziger}, \citenamefont {Koenig},\ and\ \citenamefont
  {Pines}}]{zwanziger1990non}%
  \BibitemOpen
  \bibfield  {author} {\bibinfo {author} {\bibfnamefont {J.}~\bibnamefont
  {Zwanziger}}, \bibinfo {author} {\bibfnamefont {M.}~\bibnamefont {Koenig}},\
  and\ \bibinfo {author} {\bibfnamefont {A.}~\bibnamefont {Pines}},\ }\href
  {\doibase 10.1103/PhysRevA.42.3107} {\bibfield  {journal} {\bibinfo
  {journal} {Phys. Rev. A}\ }\textbf {\bibinfo {volume} {42}},\ \bibinfo
  {pages} {3107} (\bibinfo {year} {1990})}\BibitemShut {NoStop}%
\bibitem [{\citenamefont {Zee}(1988)}]{zee1988non}%
  \BibitemOpen
  \bibfield  {author} {\bibinfo {author} {\bibfnamefont {A.}~\bibnamefont
  {Zee}},\ }\href {\doibase 10.1103/PhysRevA.38.1} {\bibfield  {journal}
  {\bibinfo  {journal} {Phys. Rev. A}\ }\textbf {\bibinfo {volume} {38}},\
  \bibinfo {pages} {1} (\bibinfo {year} {1988})}\BibitemShut {NoStop}%
\bibitem [{\citenamefont {Mead}(1987)}]{mead1987molecular}%
  \BibitemOpen
  \bibfield  {author} {\bibinfo {author} {\bibfnamefont {C.~A.}\ \bibnamefont
  {Mead}},\ }\href {\doibase 10.1103/PhysRevLett.59.161} {\bibfield  {journal}
  {\bibinfo  {journal} {Phys. Rev. Lett.}\ }\textbf {\bibinfo {volume} {59}},\
  \bibinfo {pages} {161} (\bibinfo {year} {1987})}\BibitemShut {NoStop}%
\bibitem [{\citenamefont {Mead}(1992)}]{mead1992geometric}%
  \BibitemOpen
  \bibfield  {author} {\bibinfo {author} {\bibfnamefont {C.~A.}\ \bibnamefont
  {Mead}},\ }\href {\doibase 10.1103/RevModPhys.64.51} {\bibfield  {journal}
  {\bibinfo  {journal} {Rev. Mod. Phys.}\ }\textbf {\bibinfo {volume} {64}},\
  \bibinfo {pages} {51} (\bibinfo {year} {1992})}\BibitemShut {NoStop}%
\bibitem [{\citenamefont {Abdumalikov~Jr}\ \emph {et~al.}(2013)\citenamefont
  {Abdumalikov~Jr}, \citenamefont {Fink}, \citenamefont {Juliusson},
  \citenamefont {Pechal}, \citenamefont {Berger}, \citenamefont {Wallraff},\
  and\ \citenamefont {Filipp}}]{abdumalikov2013experimental}%
  \BibitemOpen
  \bibfield  {author} {\bibinfo {author} {\bibfnamefont {A.~A.}\ \bibnamefont
  {Abdumalikov~Jr}}, \bibinfo {author} {\bibfnamefont {J.~M.}\ \bibnamefont
  {Fink}}, \bibinfo {author} {\bibfnamefont {K.}~\bibnamefont {Juliusson}},
  \bibinfo {author} {\bibfnamefont {M.}~\bibnamefont {Pechal}}, \bibinfo
  {author} {\bibfnamefont {S.}~\bibnamefont {Berger}}, \bibinfo {author}
  {\bibfnamefont {A.}~\bibnamefont {Wallraff}},\ and\ \bibinfo {author}
  {\bibfnamefont {S.}~\bibnamefont {Filipp}},\ }\href {\doibase
  10.1038/nature12010} {\bibfield  {journal} {\bibinfo  {journal} {Nature}\
  }\textbf {\bibinfo {volume} {496}},\ \bibinfo {pages} {482} (\bibinfo {year}
  {2013})}\BibitemShut {NoStop}%
\bibitem [{\citenamefont {Sugawa}\ \emph {et~al.}(2018)\citenamefont {Sugawa},
  \citenamefont {Salces-Carcoba}, \citenamefont {Perry}, \citenamefont {Yue},\
  and\ \citenamefont {Spielman}}]{sugawa2018second}%
  \BibitemOpen
  \bibfield  {author} {\bibinfo {author} {\bibfnamefont {S.}~\bibnamefont
  {Sugawa}}, \bibinfo {author} {\bibfnamefont {F.}~\bibnamefont
  {Salces-Carcoba}}, \bibinfo {author} {\bibfnamefont {A.~R.}\ \bibnamefont
  {Perry}}, \bibinfo {author} {\bibfnamefont {Y.}~\bibnamefont {Yue}},\ and\
  \bibinfo {author} {\bibfnamefont {I.}~\bibnamefont {Spielman}},\ }\href
  {\doibase 10.1126/science.aam9031} {\bibfield  {journal} {\bibinfo  {journal}
  {Science}\ }\textbf {\bibinfo {volume} {360}},\ \bibinfo {pages} {1429}
  (\bibinfo {year} {2018})}\BibitemShut {NoStop}%
\bibitem [{\citenamefont {Wu}\ and\ \citenamefont
  {Yang}(1975)}]{wu1975concept}%
  \BibitemOpen
  \bibfield  {author} {\bibinfo {author} {\bibfnamefont {T.~T.}\ \bibnamefont
  {Wu}}\ and\ \bibinfo {author} {\bibfnamefont {C.~N.}\ \bibnamefont {Yang}},\
  }\href {\doibase 10.1103/PhysRevD.12.3845} {\bibfield  {journal} {\bibinfo
  {journal} {Phys. Rev. D}\ }\textbf {\bibinfo {volume} {12}},\ \bibinfo
  {pages} {3845} (\bibinfo {year} {1975})}\BibitemShut {NoStop}%
\bibitem [{\citenamefont {Horv{\'a}thy}(1986)}]{horvathy1986non}%
  \BibitemOpen
  \bibfield  {author} {\bibinfo {author} {\bibfnamefont {P.}~\bibnamefont
  {Horv{\'a}thy}},\ }\href {\doibase 10.1103/PhysRevD.33.407} {\bibfield
  {journal} {\bibinfo  {journal} {Phys. Rev. D}\ }\textbf {\bibinfo {volume}
  {33}},\ \bibinfo {pages} {407} (\bibinfo {year} {1986})}\BibitemShut
  {NoStop}%
\bibitem [{\citenamefont {Alford}\ \emph {et~al.}(1990)\citenamefont {Alford},
  \citenamefont {March-Russell},\ and\ \citenamefont
  {Wilczek}}]{alford1990discrete}%
  \BibitemOpen
  \bibfield  {author} {\bibinfo {author} {\bibfnamefont {M.~G.}\ \bibnamefont
  {Alford}}, \bibinfo {author} {\bibfnamefont {J.}~\bibnamefont
  {March-Russell}},\ and\ \bibinfo {author} {\bibfnamefont {F.}~\bibnamefont
  {Wilczek}},\ }\href {\doibase 10.1016/0550-3213(90)90512-C} {\bibfield
  {journal} {\bibinfo  {journal} {Nuclear Physics B}\ }\textbf {\bibinfo
  {volume} {337}},\ \bibinfo {pages} {695} (\bibinfo {year}
  {1990})}\BibitemShut {NoStop}%
\bibitem [{\citenamefont {Chaichian}\ \emph {et~al.}(2002)\citenamefont
  {Chaichian}, \citenamefont {Pre{\v{s}}najder}, \citenamefont
  {Sheikh-Jabbari},\ and\ \citenamefont {Tureanu}}]{chaichian2002aharonov}%
  \BibitemOpen
  \bibfield  {author} {\bibinfo {author} {\bibfnamefont {M.}~\bibnamefont
  {Chaichian}}, \bibinfo {author} {\bibfnamefont {P.}~\bibnamefont
  {Pre{\v{s}}najder}}, \bibinfo {author} {\bibfnamefont {M.}~\bibnamefont
  {Sheikh-Jabbari}},\ and\ \bibinfo {author} {\bibfnamefont {A.}~\bibnamefont
  {Tureanu}},\ }\href {\doibase 10.1016/S0370-2693(02)01176-0} {\bibfield
  {journal} {\bibinfo  {journal} {Physics Letters B}\ }\textbf {\bibinfo
  {volume} {527}},\ \bibinfo {pages} {149} (\bibinfo {year}
  {2002})}\BibitemShut {NoStop}%
\bibitem [{\citenamefont {Osterloh}\ \emph {et~al.}(2005)\citenamefont
  {Osterloh}, \citenamefont {Baig}, \citenamefont {Santos}, \citenamefont
  {Zoller},\ and\ \citenamefont {Lewenstein}}]{osterloh2005cold}%
  \BibitemOpen
  \bibfield  {author} {\bibinfo {author} {\bibfnamefont {K.}~\bibnamefont
  {Osterloh}}, \bibinfo {author} {\bibfnamefont {M.}~\bibnamefont {Baig}},
  \bibinfo {author} {\bibfnamefont {L.}~\bibnamefont {Santos}}, \bibinfo
  {author} {\bibfnamefont {P.}~\bibnamefont {Zoller}},\ and\ \bibinfo {author}
  {\bibfnamefont {M.}~\bibnamefont {Lewenstein}},\ }\href {\doibase
  10.1103/PhysRevLett.95.010403} {\bibfield  {journal} {\bibinfo  {journal}
  {Phys. Rev. Lett.}\ }\textbf {\bibinfo {volume} {95}},\ \bibinfo {pages}
  {010403} (\bibinfo {year} {2005})}\BibitemShut {NoStop}%
\bibitem [{\citenamefont {Jacob}\ \emph {et~al.}(2007)\citenamefont {Jacob},
  \citenamefont {{\"O}hberg}, \citenamefont {Juzeli{\=u}nas},\ and\
  \citenamefont {Santos}}]{jacob2007cold}%
  \BibitemOpen
  \bibfield  {author} {\bibinfo {author} {\bibfnamefont {A.}~\bibnamefont
  {Jacob}}, \bibinfo {author} {\bibfnamefont {P.}~\bibnamefont {{\"O}hberg}},
  \bibinfo {author} {\bibfnamefont {G.}~\bibnamefont {Juzeli{\=u}nas}},\ and\
  \bibinfo {author} {\bibfnamefont {L.}~\bibnamefont {Santos}},\ }\href
  {\doibase 10.1007/s00340-007-2865-6} {\bibfield  {journal} {\bibinfo
  {journal} {Appl. Phys. B}\ }\textbf {\bibinfo {volume} {89}},\ \bibinfo
  {pages} {439} (\bibinfo {year} {2007})}\BibitemShut {NoStop}%
\bibitem [{\citenamefont {Iadecola}\ \emph {et~al.}(2016)\citenamefont
  {Iadecola}, \citenamefont {Schuster},\ and\ \citenamefont
  {Chamon}}]{iadecola2016non}%
  \BibitemOpen
  \bibfield  {author} {\bibinfo {author} {\bibfnamefont {T.}~\bibnamefont
  {Iadecola}}, \bibinfo {author} {\bibfnamefont {T.}~\bibnamefont {Schuster}},\
  and\ \bibinfo {author} {\bibfnamefont {C.}~\bibnamefont {Chamon}},\ }\href
  {\doibase 10.1103/PhysRevLett.117.073901} {\bibfield  {journal} {\bibinfo
  {journal} {Phys. Rev. Lett.}\ }\textbf {\bibinfo {volume} {117}},\ \bibinfo
  {pages} {073901} (\bibinfo {year} {2016})}\BibitemShut {NoStop}%
\bibitem [{\citenamefont {Chen}\ \emph {et~al.}(2018)\citenamefont {Chen},
  \citenamefont {Zhang}, \citenamefont {Xiong}, \citenamefont {Shen},\ and\
  \citenamefont {Chan}}]{chen2018non}%
  \BibitemOpen
  \bibfield  {author} {\bibinfo {author} {\bibfnamefont {Y.}~\bibnamefont
  {Chen}}, \bibinfo {author} {\bibfnamefont {R.-Y.}\ \bibnamefont {Zhang}},
  \bibinfo {author} {\bibfnamefont {Z.}~\bibnamefont {Xiong}}, \bibinfo
  {author} {\bibfnamefont {J.~Q.}\ \bibnamefont {Shen}},\ and\ \bibinfo
  {author} {\bibfnamefont {C.}~\bibnamefont {Chan}},\ }\href@noop {} {\bibfield
   {journal} {\bibinfo  {journal} {arXiv preprint arXiv:1802.09866}\ }
  (\bibinfo {year} {2018})}\BibitemShut {NoStop}%
\bibitem [{\citenamefont {Bohm}\ \emph {et~al.}(2013)\citenamefont {Bohm},
  \citenamefont {Mostafazadeh}, \citenamefont {Koizumi}, \citenamefont {Niu},\
  and\ \citenamefont {Zwanziger}}]{bohm2013geometric}%
  \BibitemOpen
  \bibfield  {author} {\bibinfo {author} {\bibfnamefont {A.}~\bibnamefont
  {Bohm}}, \bibinfo {author} {\bibfnamefont {A.}~\bibnamefont {Mostafazadeh}},
  \bibinfo {author} {\bibfnamefont {H.}~\bibnamefont {Koizumi}}, \bibinfo
  {author} {\bibfnamefont {Q.}~\bibnamefont {Niu}},\ and\ \bibinfo {author}
  {\bibfnamefont {J.}~\bibnamefont {Zwanziger}},\ }\href@noop {} {\emph
  {\bibinfo {title} {The Geometric Phase in Quantum Systems: Foundations,
  Mathematical Concepts, and Applications in Molecular and Condensed Matter
  Physics}}}\ (\bibinfo  {publisher} {Springer Science \& Business Media},\
  \bibinfo {year} {2013})\BibitemShut {NoStop}%
\bibitem [{\citenamefont {Wilson}(1974)}]{wilson1974confinement}%
  \BibitemOpen
  \bibfield  {author} {\bibinfo {author} {\bibfnamefont {K.~G.}\ \bibnamefont
  {Wilson}},\ }\href {\doibase 10.1103/PhysRevD.10.2445} {\bibfield  {journal}
  {\bibinfo  {journal} {Phys. Rev. D}\ }\textbf {\bibinfo {volume} {10}},\
  \bibinfo {pages} {2445} (\bibinfo {year} {1974})}\BibitemShut {NoStop}%
\bibitem [{\citenamefont {Goldman}\ \emph {et~al.}(2009)\citenamefont
  {Goldman}, \citenamefont {Kubasiak}, \citenamefont {Gaspard},\ and\
  \citenamefont {Lewenstein}}]{goldman2009ultracold}%
  \BibitemOpen
  \bibfield  {author} {\bibinfo {author} {\bibfnamefont {N.}~\bibnamefont
  {Goldman}}, \bibinfo {author} {\bibfnamefont {A.}~\bibnamefont {Kubasiak}},
  \bibinfo {author} {\bibfnamefont {P.}~\bibnamefont {Gaspard}},\ and\ \bibinfo
  {author} {\bibfnamefont {M.}~\bibnamefont {Lewenstein}},\ }\href {\doibase
  10.1103/PhysRevA.79.023624} {\bibfield  {journal} {\bibinfo  {journal} {Phys.
  Rev. A}\ }\textbf {\bibinfo {volume} {79}},\ \bibinfo {pages} {023624}
  (\bibinfo {year} {2009})}\BibitemShut {NoStop}%
\bibitem [{\citenamefont {Benalcazar}\ \emph {et~al.}(2017)\citenamefont
  {Benalcazar}, \citenamefont {Bernevig},\ and\ \citenamefont
  {Hughes}}]{benalcazar2017quantized}%
  \BibitemOpen
  \bibfield  {author} {\bibinfo {author} {\bibfnamefont {W.~A.}\ \bibnamefont
  {Benalcazar}}, \bibinfo {author} {\bibfnamefont {B.~A.}\ \bibnamefont
  {Bernevig}},\ and\ \bibinfo {author} {\bibfnamefont {T.~L.}\ \bibnamefont
  {Hughes}},\ }\href {\doibase 10.1126/science.aah6442} {\bibfield  {journal}
  {\bibinfo  {journal} {Science}\ }\textbf {\bibinfo {volume} {357}},\ \bibinfo
  {pages} {61} (\bibinfo {year} {2017})}\BibitemShut {NoStop}%
\bibitem [{\citenamefont {Peterson}\ \emph {et~al.}(2018)\citenamefont
  {Peterson}, \citenamefont {Benalcazar}, \citenamefont {Hughes},\ and\
  \citenamefont {Bahl}}]{peterson2018quantized}%
  \BibitemOpen
  \bibfield  {author} {\bibinfo {author} {\bibfnamefont {C.~W.}\ \bibnamefont
  {Peterson}}, \bibinfo {author} {\bibfnamefont {W.~A.}\ \bibnamefont
  {Benalcazar}}, \bibinfo {author} {\bibfnamefont {T.~L.}\ \bibnamefont
  {Hughes}},\ and\ \bibinfo {author} {\bibfnamefont {G.}~\bibnamefont {Bahl}},\
  }\href {\doibase 10.1038/nature25777} {\bibfield  {journal} {\bibinfo
  {journal} {Nature}\ }\textbf {\bibinfo {volume} {555}},\ \bibinfo {pages}
  {346} (\bibinfo {year} {2018})}\BibitemShut {NoStop}%
\bibitem [{\citenamefont {Serra-Garcia}\ \emph {et~al.}(2018)\citenamefont
  {Serra-Garcia}, \citenamefont {Peri}, \citenamefont {S{\"u}sstrunk},
  \citenamefont {Bilal}, \citenamefont {Larsen}, \citenamefont {Villanueva},\
  and\ \citenamefont {Huber}}]{serra2018observation}%
  \BibitemOpen
  \bibfield  {author} {\bibinfo {author} {\bibfnamefont {M.}~\bibnamefont
  {Serra-Garcia}}, \bibinfo {author} {\bibfnamefont {V.}~\bibnamefont {Peri}},
  \bibinfo {author} {\bibfnamefont {R.}~\bibnamefont {S{\"u}sstrunk}}, \bibinfo
  {author} {\bibfnamefont {O.~R.}\ \bibnamefont {Bilal}}, \bibinfo {author}
  {\bibfnamefont {T.}~\bibnamefont {Larsen}}, \bibinfo {author} {\bibfnamefont
  {L.~G.}\ \bibnamefont {Villanueva}},\ and\ \bibinfo {author} {\bibfnamefont
  {S.~D.}\ \bibnamefont {Huber}},\ }\href {\doibase 10.1038/nature25156}
  {\bibfield  {journal} {\bibinfo  {journal} {Nature}\ }\textbf {\bibinfo
  {volume} {555}},\ \bibinfo {pages} {342} (\bibinfo {year}
  {2018})}\BibitemShut {NoStop}%
\bibitem [{\citenamefont {Zilberberg}\ \emph {et~al.}(2018)\citenamefont
  {Zilberberg}, \citenamefont {Huang}, \citenamefont {Guglielmon},
  \citenamefont {Wang}, \citenamefont {Chen}, \citenamefont {Kraus},\ and\
  \citenamefont {Rechtsman}}]{zilberberg2018photonic}%
  \BibitemOpen
  \bibfield  {author} {\bibinfo {author} {\bibfnamefont {O.}~\bibnamefont
  {Zilberberg}}, \bibinfo {author} {\bibfnamefont {S.}~\bibnamefont {Huang}},
  \bibinfo {author} {\bibfnamefont {J.}~\bibnamefont {Guglielmon}}, \bibinfo
  {author} {\bibfnamefont {M.}~\bibnamefont {Wang}}, \bibinfo {author}
  {\bibfnamefont {K.~P.}\ \bibnamefont {Chen}}, \bibinfo {author}
  {\bibfnamefont {Y.~E.}\ \bibnamefont {Kraus}},\ and\ \bibinfo {author}
  {\bibfnamefont {M.~C.}\ \bibnamefont {Rechtsman}},\ }\href {\doibase
  10.1038/nature25011} {\bibfield  {journal} {\bibinfo  {journal} {Nature}\
  }\textbf {\bibinfo {volume} {553}},\ \bibinfo {pages} {59} (\bibinfo {year}
  {2018})}\BibitemShut {NoStop}%
\bibitem [{\citenamefont {Lohse}\ \emph {et~al.}(2018)\citenamefont {Lohse},
  \citenamefont {Schweizer}, \citenamefont {Price}, \citenamefont
  {Zilberberg},\ and\ \citenamefont {Bloch}}]{lohse2018exploring}%
  \BibitemOpen
  \bibfield  {author} {\bibinfo {author} {\bibfnamefont {M.}~\bibnamefont
  {Lohse}}, \bibinfo {author} {\bibfnamefont {C.}~\bibnamefont {Schweizer}},
  \bibinfo {author} {\bibfnamefont {H.~M.}\ \bibnamefont {Price}}, \bibinfo
  {author} {\bibfnamefont {O.}~\bibnamefont {Zilberberg}},\ and\ \bibinfo
  {author} {\bibfnamefont {I.}~\bibnamefont {Bloch}},\ }\href {\doibase
  10.1038/nature25000} {\bibfield  {journal} {\bibinfo  {journal} {Nature}\
  }\textbf {\bibinfo {volume} {553}},\ \bibinfo {pages} {55} (\bibinfo {year}
  {2018})}\BibitemShut {NoStop}%
\bibitem [{\citenamefont {Wang}\ \emph {et~al.}(2018)\citenamefont {Wang},
  \citenamefont {Zhang}, \citenamefont {Chen}, \citenamefont {Bertrand},
  \citenamefont {Shams-Ansari}, \citenamefont {Chandrasekhar}, \citenamefont
  {Winzer},\ and\ \citenamefont {Lon{\v{c}}ar}}]{wang2018integrated}%
  \BibitemOpen
  \bibfield  {author} {\bibinfo {author} {\bibfnamefont {C.}~\bibnamefont
  {Wang}}, \bibinfo {author} {\bibfnamefont {M.}~\bibnamefont {Zhang}},
  \bibinfo {author} {\bibfnamefont {X.}~\bibnamefont {Chen}}, \bibinfo {author}
  {\bibfnamefont {M.}~\bibnamefont {Bertrand}}, \bibinfo {author}
  {\bibfnamefont {A.}~\bibnamefont {Shams-Ansari}}, \bibinfo {author}
  {\bibfnamefont {S.}~\bibnamefont {Chandrasekhar}}, \bibinfo {author}
  {\bibfnamefont {P.}~\bibnamefont {Winzer}},\ and\ \bibinfo {author}
  {\bibfnamefont {M.}~\bibnamefont {Lon{\v{c}}ar}},\ }\href {\doibase
  10.1038/s41586-018-0551-y} {\bibfield  {journal} {\bibinfo  {journal}
  {Nature}\ }\textbf {\bibinfo {volume} {562}},\ \bibinfo {pages} {101}
  (\bibinfo {year} {2018})}\BibitemShut {NoStop}%
\bibitem [{\citenamefont {Bi}(2018)}]{bi2018materials}%
  \BibitemOpen
  \bibfield  {author} {\bibinfo {author} {\bibfnamefont {L.}~\bibnamefont
  {Bi}},\ }\href {\doibase 10.1557/mrs.2018.120} {\bibfield  {journal}
  {\bibinfo  {journal} {MRS Bulletin}\ }\textbf {\bibinfo {volume} {43}},\
  \bibinfo {pages} {408} (\bibinfo {year} {2018})}\BibitemShut {NoStop}%
\bibitem [{\citenamefont {Burrello}\ and\ \citenamefont
  {Trombettoni}(2010)}]{burrello2010non}%
  \BibitemOpen
  \bibfield  {author} {\bibinfo {author} {\bibfnamefont {M.}~\bibnamefont
  {Burrello}}\ and\ \bibinfo {author} {\bibfnamefont {A.}~\bibnamefont
  {Trombettoni}},\ }\href {\doibase 10.1103/PhysRevLett.105.125304} {\bibfield
  {journal} {\bibinfo  {journal} {Phys. Rev. Lett.}\ }\textbf {\bibinfo
  {volume} {105}},\ \bibinfo {pages} {125304} (\bibinfo {year}
  {2010})}\BibitemShut {NoStop}%
\bibitem [{\citenamefont {Keilmann}\ \emph {et~al.}(2011)\citenamefont
  {Keilmann}, \citenamefont {Lanzmich}, \citenamefont {McCulloch},\ and\
  \citenamefont {Roncaglia}}]{keilmann2011statistically}%
  \BibitemOpen
  \bibfield  {author} {\bibinfo {author} {\bibfnamefont {T.}~\bibnamefont
  {Keilmann}}, \bibinfo {author} {\bibfnamefont {S.}~\bibnamefont {Lanzmich}},
  \bibinfo {author} {\bibfnamefont {I.}~\bibnamefont {McCulloch}},\ and\
  \bibinfo {author} {\bibfnamefont {M.}~\bibnamefont {Roncaglia}},\ }\href
  {\doibase 10.1038/ncomms1353} {\bibfield  {journal} {\bibinfo  {journal}
  {Nat. Commun.}\ }\textbf {\bibinfo {volume} {2}},\ \bibinfo {pages} {361}
  (\bibinfo {year} {2011})}\BibitemShut {NoStop}%
\bibitem [{\citenamefont {Yuan}\ \emph {et~al.}(2017)\citenamefont {Yuan},
  \citenamefont {Xiao}, \citenamefont {Xu},\ and\ \citenamefont
  {Fan}}]{yuan2017creating}%
  \BibitemOpen
  \bibfield  {author} {\bibinfo {author} {\bibfnamefont {L.}~\bibnamefont
  {Yuan}}, \bibinfo {author} {\bibfnamefont {M.}~\bibnamefont {Xiao}}, \bibinfo
  {author} {\bibfnamefont {S.}~\bibnamefont {Xu}},\ and\ \bibinfo {author}
  {\bibfnamefont {S.}~\bibnamefont {Fan}},\ }\href {\doibase
  10.1103/PhysRevA.96.043864} {\bibfield  {journal} {\bibinfo  {journal} {Phys.
  Rev. A}\ }\textbf {\bibinfo {volume} {96}},\ \bibinfo {pages} {043864}
  (\bibinfo {year} {2017})}\BibitemShut {NoStop}%
\end{thebibliography}

\section*{Acknowledgements}
    We thank Yifan Lin, Rongya Luo, and Shouzhi Yang for assistance in the setup and measurement.
    We thank fruitful discussions with Karl Berggren, Dirk Englund, Liang Fu, Morten Kjaergaard, Junru Li, Eugene J. Mele, William D. Oliver, Ren-Jye Shiue, and Ashvin Vishwanath.
	We thank Paola Rebusco and Jamison Sloan for reading and editing of the manuscript.
	Research was sponsored in part by the Army Research Office and was accomplished under Cooperative Agreement Number W911NF-18-2-0048.
    This material is based upon work supported in part by the National Science Foundation under Grant No. CCF-1640012.

\end{document}